\documentclass[aps,12pt,tightenlines,amsmath,amssymb]{revtex4}

\DeclareMathOperator{\tr}{{\rm Tr}}
\DeclareMathOperator{\re}{{\rm Re}}
\newcommand{\U}{^{(U)}}
\newcommand{\Ut}{^{(U(t))}}
\newcommand{\du}{\delta\U}
\newcommand{\akn}{a_k^{(n)}}
\newcommand{\takn}{\tilde{a}_k^{(n)}}

\newcommand{\taln}{\tilde{a}_l^{(n)}}
\newcommand{\talnc}{\tilde{a}_l^{(n)*}}
\newcommand{\tann}{\tilde{a}_n^{(n)}}
\newcommand{\tannc}{\tilde{a}_n^{(n)*}}
\newcommand{\eno}{E_n^{(0)}}
\newcommand{\eko}{E_k^{(0)}}
\newcommand{\elo}{E_l^{(0)}}
\newcommand{\ud}{{\rm d}}
\newcommand{\kpno}{|\psi_n^{(0)}\rangle}
\newcommand{\kpko}{|\psi_k^{(0)}\rangle}

\newcommand{\bpno}{\langle\psi_n^{(0)}|}

\newcommand{\bplo}{\langle\psi_l^{(0)}|}
\newcommand{\kak}{|\psi_{A,k}^{(0)}\rangle}
\newcommand{\bak}{\langle\psi_{A,k}^{(0)}|}
\newcommand{\kbl}{|\psi_{B,l}^{(0)}\rangle}
\newcommand{\bbl}{\langle\psi_{B,l}^{(0)}|}
\newcommand{\eako}{E_{A,k}^{(0)}}
\newcommand{\eano}{E_{A,n}^{(0)}}
\newcommand{\eblo}{E_{B,l}^{(0)}}
\newcommand{\ebmo}{E_{B,m}^{(0)}}
\newcommand{\eak}{E_{A,k}}
\newcommand{\ean}{E_{A,n}}
\newcommand{\ebl}{E_{B,l}}
\newcommand{\ebm}{E_{B,m}}
\newcommand{\taklnm}{\tilde{a}_{(k,l)}^{(n,m)}}
\newcommand{\vklnm}{V_{(k,l)}^{(n,m)}}

\def\Tr{\mathop{\mathrm{Tr}}\nolimits}
\def\Eqref#1{Eq.~(\ref{#1})}
\def\Eqsref#1#2{Eqs.~(\ref{#1}) and (\ref{#2})}
\def\qqquad{\qquad\qquad}
\def\qqqquad{\qqquad\qqquad}
\def\qqqqquad{\qqqquad\qqqquad}

\newcommand{\be}{\begin{equation}}\newcommand{\ee}{\end{equation}}
\newcommand{\bea}{\begin{eqnarray}}\newcommand{\eea}{\end{eqnarray}}
\newcommand{\bean}{\begin{eqnarray*}}\newcommand{\eean}{\end{eqnarray*}}

\newcounter{remark}

\def\dbar{{\,\mathchar'26\mkern-12mu d}}

\def\AplusB{$A\mskip-2mu+B$}

\begin{document}

\title{Relative entropy, interaction energy and the nature of dissipation}

\author{B. Gaveau}
\affiliation{Laboratoire analyse et physique math\'ematique, 14 avenue F\'elix Faure, 75015 Paris, France}

\author{L. Granger}
\affiliation{Max Planck Institute for the Physics of Complex Systems, N\"othnitzer Str.\ 38, D-01187 Dresden, Germany}

\author{M. Moreau}
\affiliation{Universit\'e Pierre et Marie Curie, LPTMC, case 121, 4 pl.\ Jussieu, 75252 Paris cedex 05, France}

\author{L. S. Schulman}
\affiliation{Physics Department, Clarkson University, Potsdam, New York 13699-5820, USA}
\email{schulman@clarkson.edu}

\date{\today}
\begin{abstract}
Many thermodynamic relations involve inequalities, with equality if a process does not involve dissipation. In this article we provide equalities in which the dissipative contribution is shown to involve the relative entropy (a.k.a.\ Kullback-Leibler divergence). The processes considered are general time evolutions both in classical and quantum mechanics, and the initial state is sometimes thermal, sometimes partially so. By calculating a transport coefficient we show that indeed---at least in this case---the source of dissipation in that coefficient is the relative entropy.
\end{abstract}

\maketitle

\section{Introduction \label{s.intro}}

The distinction between heat and work, between the uncontrollable flow of energy of molecular processes and the controllable flow of energy usable by an agent, underlies all of thermodynamics, and is implicitly incorporated in the equation $dE=\dbar W+\dbar Q$\@. This distinction is defined by human subjectivity and by the human technological ability to extract work from the flow of energy of microscopic processes.

On the other hand, assuming that the evolution is given by the fundamental laws of dynamics, classical or quantum, one must represent the evolution of a system at a fundamental microscopic level by the action of a unitary operator on a quantum state (density operator) in quantum situations or by the action of a symplectic operator on a classical state (probability distribution) in classical situations. The state of the system evolves according to the Heisenberg equation of motion or according to the Liouville equation respectively. An immediate consequence is that \textit{the entropy of the exact microscopic state of a system that is isolated or is coupled only to an external source of work stays constant during the evolution. In particular, the microscopic state cannot tend to an equilibrium state}. Thus, the information content of the exact microscopic state stays constant during the evolution. The problem is that in practice it is impossible both to define the microscopic state and to follow its exact evolution. The definition of the state and the representation of the exact evolution using unitary or symplectic dynamics are thus untenable idealizations. Nevertheless, and this constitutes a paradox, these idealizations cannot be ignored or dismissed, because it is precisely the difference between the exact evolution and its standard approximations which explains and can be used to predict dissipative effects, both of energy and information.

In thermodynamics, in kinetic theories or in stochastic dynamics, the exact microscopic state of a system is replaced by an approximate or ``coarse-grained'' state and the corresponding exact evolution is replaced by an evolution of the corresponding coarse-grained state (or, in standard thermodynamics by a quasi-static or formal evolution). There are two main reasons for using these approximate states and evolutions: \\
1. As discussed, it is impossible---even in principle---to specify the exact state of a large system and follow its evolution. An attempt at extremely high precision would modify the system, even in a classical context (related to Maxwell's demon). And it is even worse for quantum systems. Moreover, this would be useless.\\
2. Only slow variables (on the time scales of microscopic processes) can be measured with confidence and stability. As a result, an observer can only describe the system as a state of minimal information (or maximal entropy) compatible with the observed slow variables~\cite{landau, grains}.

The coarse-grained state is thus a statistical data structure which summarizes at a given moment the knowledge of the observer. The evolution of this coarse-grained state merely reflects the evolution of the knowledge of the observer about the system. The observer cannot follow the microscopic processes, but only the slow variables which can be measured and used, and as a consequence there is a loss of information about the details of the microscopic processes; in the traditional language of thermodynamics, entropy increases or is produced. The observer, reflecting a particular state of knowledge (or more precisely, a lack of knowledge), describes the state of the system as a state of minimal information (or maximal entropy) compatible with observation. Thus, entropy is not a kind of substance flowing from one part of a system to another part or mysteriously produced internally by the physical system, as is often suggested by many texts of thermodynamics or statistical physics: it is only the observer's partial inability to relate the exact microscopic theory to a reduced macroscopic description in order to use the system as a source of useful work or information. This is what is measured as an increase of entropy or by entropy production. The macroscopic state is the result of a statistical inference on the observed variables (which are the slow variables of the system \cite{landau, grains}. This point of view on the nature of entropy was emphasized by Jaynes, who observed \cite{jaynes1, jaynes2}, ``The expression `irreversible process' represents a semantic confusion\dots''

The difference between the exact evolution of the microscopic ideal state and the evolution of its coarse-grained approximation is what is called ``dissipation,'' both of information content and of energy or other ``useful'' variables. Standard thermodynamics uses the maximal coarse graining of equilibrium, and the idealized evolution is not modeled explicitly, so dissipation can be taken into account only by inequalities. For more detailed coarse-graining (as in hydrodynamics, Boltzmann's equation, kinetic theories or stochastic thermodynamics) one can obtain an estimate for the dissipative effects, for example, by the calculation of transport coefficients.

In this article, our main purpose is to prove that the relative entropy term between the initial and final states measures dissipation. In our approach, ``dissipation'' is defined as the difference between the maximal work that the physicist thinks could be extracted from a system when using the thermodynamic or quasi-static theory to make predictions, and the work that is actually extracted because the system is evolving according to the exact dynamics, classical or quantum, independently of what the physicist thinks (see also our use of relative entropy in \cite{master} and \cite{framework}, where the context was more limited \cite{note:limited}). Moreover, in the present context we find that the relative entropy terms are proportional to the square of the interaction energy. In all standard theories, dissipative effects are measured by the transport coefficients of energy or momentum or concentration of chemical species. Thus, we need prove that the relative entropy allows the calculation of transport coefficients. Indeed, we show below that the relative entropy terms provide the calculation of the thermal conductivity between two general quantum systems, initially at thermal equilibrium at different temperatures. This is a kind of Fourier law, except that we do not suppose a linear regime, so that the temperature dependence is more complicated than simple linearity. Moreover, our exact calculation of the transport coefficient shows that it is indeed proportional to the square of the interaction energy, which confirms that for vanishingly small interaction energy no transfer occurs in finite time. In other words, no power or finite rate of information flow can be extracted from a system if one does not have at the same time dissipative effects.

In the following material, we first consider a system comprised of two components, $A$ and $B$\@. We make no specific hypotheses on the size of the systems, and we do not introduce thermal reservoirs. Thus, the identities we derive are in effect exact tautologies. In Secs.\ \ref{s.bg3} to \ref{s.bg5}, we present several identities. We here mention two examples: 1)~a derivation of the Brillouin-Landauer estimate of the energy necessary to change the information content of a system; 2)~an estimate of the work that can be extracted from a two-part system in interaction with an external source of work in terms of non-equilibrium free energies and relative entropy of the state before and after the evolution. Similar identities were also obtained recently by Esposito et al. \cite{esposito1}, Reeb and Wolf \cite{reeb}, and Takara et al.~\cite{takara} Continuing, we study the effect of an external agent on an (otherwise) isolated system; again we obtain an identity relating the work to the difference of internal (not the free) energies along with the usual dissipative terms. Then, we derive the relation between the relative entropy and the heat conductivity in a quantum system. Finally, we define a general notion of coarse-graining or reduced description, which includes the usual notions. In some of our examples one or both systems are initially at thermal equilibrium, but only the initial temperatures appear explicitly in the definition of the non-equilibrium free energies. The latter are no longer state functions because they depend explicitly of the initial temperature and not of the actual effective temperature. No coarse graining by an effective final or intermediate thermal state is used, and neither system is a reservoir.

\section{Notations and basic identities\label{s.bg2}}

\subsection{States and entropy\label{s.bg2.1}}

Many results will be valid both in classical and quantum contexts. We denote by $\rho$ either a probability distribution function over a classical phase space, or a density matrix in the quantum case. We denote by $\tr$ either the integral on the phase space, or the trace operation. Thus $\rho$ is a positive quantity and satisfies $\tr \rho = 1$\@. The entropy of $\rho$ is
\begin{equation}
	S(\rho) = - \tr \rho \log \rho
\label{e.bg2.1}
\,.
\end{equation}
It is defined up to a multiplicative constant. (Classically $\rho$ should be divided by a dimensional constant to render it dimensionless.)

The relative entropy (see \cite{cover}) is defined by
\begin{equation}
	S(\rho|\rho') = \tr \left( \rho\left( \log \rho - \log \rho'
	\right) \right)
\,,
\label{e.bg2.2}
\end{equation}
where $\rho$ and $\rho'$ are states.

One has
\begin{equation}
	S(\rho|\rho') \ge 0
\,,
\label{e.bg2.3}
\end{equation}
and $S(\rho|\rho')$ does not depend on the units in phase space. Moreover $S(\rho|\rho') = 0$ if and only if $\rho = \rho'$\@.

Writing $S(\rho|\rho')$ as $-\tr\rho\log\rho'-(-\tr \rho\log\rho)$, suggests the following interpretation:  Suppose the true state is $\rho$, but the observer thinks that the state is $\rho'$\@.  $S(\rho|\rho')$ is then the true average of the missing information versus the estimate of the missing information.

\subsection{The basic identity\label{s.bg2.2}}

If we add and subtract $S(\rho')$ in the second member of \Eqref{e.bg2.2}, we obtain the basic identity
\begin{equation}
	S(\rho|\rho') = S(\rho') - S(\rho) - \tr\left( (\rho-\rho')
	\log \rho'
	\right)
\label{e.bg2.4}
\,.
\end{equation}
Most of our results follow from this identity.

When $\rho'$ is a thermal state,
\begin{equation}
\rho'=\rho_\beta = \frac{e^{-\beta H}}{Z(\beta,H)}
\,,
\label{e.bg2.5}
\end{equation}
where
\begin{equation}
	Z(\beta,H) = \tr e^{-\beta H}
\label{e.bg2.5a}
\end{equation}
is the partition function. With $\rho'=\rho_\beta$, the identity (\ref{e.bg2.4}) reduces to
\begin{equation}
	S(\rho|\rho_\beta) = S(\rho_\beta) - S(\rho) + \beta \tr\left((\rho-\rho_\beta)H \right)
\,.
\label{e.bg2.6}
\end{equation}
Here $H$ is a given function or operator.

Defining the free energy of state $\rho$ by
\begin{equation}
	F(\rho,H) = \tr\left( \rho H \right) - \frac{1}{\beta}S(\rho)
\,,
\label{e.bg2.7}
\end{equation}
we obtain
\begin{equation}
	S(\rho|\rho_\beta) = \beta\left( F(\rho,H) - F(\rho_\beta,H) \right)
\,,
\label{e.bg2.8}
\end{equation}
and $F(\rho_\beta,H)$ is the equilibrium free energy related to the partition function by
\begin{equation}
	Z(\beta,H) = \exp\left( -\beta F(\rho_\beta,H) \right)
\,.
\label{e.bg2.9}
\end{equation}

\subsection{Evolution operators and entropy\label{s.bg2.3}}

We assume that the system (classical or quantum) evolves under the action of an arbitrary operator $U$ (symplectic  or unitary). If $\rho$ is a state, we denote by $\rho\U$ the new state after the evolution $U$\@.

Entropy is conserved by the evolution
\begin{equation}
	S(\rho\U) = S(\rho).
\label{e.bg2.10}
\end{equation}
For example in the quantum case, we have $\rho\U=U\rho\, U^\dagger$, where $U$ is the propagator: $i\frac{dU}{dt}=[H,U]$, $U|_{t=0}=1$, with $H$ a possibly time-dependent Hamiltonian.

If $\phi(\rho)$ is a functional of $\rho$ which evolves with $U$, and $\phi(\rho\U)$ is the functional after evolution of $\rho$, we denote the variation of $\phi(\rho)$ after the evolution $U$ in the following way
\begin{equation}
\delta\U (\phi(\rho))= \phi(\rho\U)-\phi(\rho)
\,.
\label{e.bg2.11}
\end{equation}

\medskip
\refstepcounter{remark} \label{r.bg1122a}
\smallskip\noindent\textsf{Remark \arabic{remark}}:~
Many of our results are valid for a general evolution $U$ which is not symplectic or unitary, for example stochastic evolution.

\section{Two systems in interaction\label{s.bg3}}

\subsection{Hypotheses\label{s.bg3.1}}

We assume that the system is formed of two parts, $A$ and $B$, in interaction. At time-0, the state is a product
state
\begin{equation}
	\rho_0 = \rho_{A,0} \otimes \rho_{B,0}
\,.
\label{e.bg3.1}
\end{equation}
After the evolution $U$, the state is $\rho\U$ and we denote by $\rho_A\U$ and $\rho_B\U$ its marginals,
\begin{equation}
	\rho_A\U = \tr_B \rho\U \quad \hbox{and}\quad \rho_B\U = \tr_A \rho\U
\,,
\label{e.bg3.2}
\end{equation}
which are then states on $A$ and $B$ respectively. We also assume that there is a quantity $H$ that is conserved by the evolution and $H$ has the form
\begin{equation}
	H = H_A + H_B + V_{AB}
\,,
\label{e.bg3.3}
\end{equation}
where $H_A$ and $H_B$ are quantities depending only on $A$ and $B$ respectively and $V_{AB}$ is an interaction term. Then, if we denote
\begin{eqnarray}
	E(\rho) &=& \tr\left( \rho H \right) = E_A(\rho) + E_B(\rho) +
	E_V(\rho)
\label{e.bg3.4}\\
	E_A(\rho) &=& \tr\left( \rho H_A \right) = \tr\left( \rho_A H_A
	\right)
\label{e.bg3.5}\\
	E_B(\rho) &=& \tr\left( \rho H_B \right) = \tr\left( \rho_B H_B
	\right)
\label{e.bg3.6}\\
	E_V(\rho) &=& \tr\left( \rho V_{AB} \right)
\,,
\label{e.bg3.7}
\end{eqnarray}
our hypothesis is that
\begin{equation}
	\du E(\rho)\equiv E(\rho\U)-E(\rho_0) = 0
\,.
\label{e.bg3.8}
\end{equation}
In particular this is the case if $U$ is time-evolution with Hamiltonian $H$\@.

\refstepcounter{remark} \label{r.bg1122b}
\smallskip\noindent\textsf{Remark \arabic{remark}}:~
For this situation certain results are also valid without the assumption that the evolution $U$ preserves the energy~$H$\@.

\subsection{Relation between a state and its marginals\label{s.bg3.2}}

Assuming \Eqref{e.bg3.1} (that the initial state is a product state), one has the identity
\begin{equation}
	\du S(\rho_A) + \du S(\rho_B) = S(\rho\U|\rho_A\U \otimes
	\rho_B\U)
\,.
\label{e.bg3.9}
\end{equation}
Indeed, using the conservation of the entropy of $\rho$ during the evolution~$U$,
\bean
\du S(\rho_A)+\du S(\rho_B)&=&S(\rho_A\U) + S(\rho_B\U)-S(\rho_A\U\otimes\rho_B\U) \\
&=&S(\rho_A\U) + S(\rho_B\U)-S(\rho\U) =S(\rho\U|\rho_A\U\otimes\rho_B\U)
\,.
\eean
This is because one evidently has $-\Tr\left(\rho\U\log\rho_A\U\right)=-\Tr\left(\rho_A\U\log\rho_A\U\right)$\@. Note that \Eqref{e.bg3.9} requires that $U$ preserve the entropy. One has also the well-known inequality
\be
S(\rho\U)\le S(\rho_A\U)+S(\rho_B\U)
\,,
\label{e.bg3.10}
\ee
which is a particular case of
\be
S(\rho)\le S(\rho_A)+S(\rho_B)
\label{e.bg3.11}
\ee
for any state $\rho$\@.

\subsection{The case where $A$ is initially in a thermal state\label{s.bg3.3}}

At time 0 we take $\rho_{A,0}$ to be thermal with temperature $\beta_A$,
\begin{equation}
	\rho_{A,0} = \rho_{A,\beta_A} =
	\frac{e^{-\beta_A H_A}}{Z_A(\beta_A)}
\,,
\label{e.bg3.12}
\end{equation}
where $Z_A(\beta_A) =Z_A(\beta_A,H_A)$ is the partition function, (\ref{e.bg2.5a}). From \Eqref{e.bg2.6} with $\rho\to\rho_A\U$ and $\rho_B\to\rho_{A,\beta_A}$, we deduce
\begin{equation}
	\du S(\rho_A) - \beta_A \du E_A(\rho_A) = -
	S(\rho_A\U|\rho_{A,\beta_A})
\,,
\label{e.bg3.13}
\end{equation}
and as a consequence
\begin{equation}
	\du S(\rho_A) - \beta_A \du E_A(\rho_A) \le 0
\,.
\label{e.bg3.14}
\end{equation}
The last two equations do not require that $U$ be a unitary evolution conserving the entropy, nor that it conserve the energy.

\refstepcounter{remark} \label{r.bg1}
\smallskip\noindent\textsf{Remark \arabic{remark}}:~
This inequality can be found in \cite{partovi} as an unnumbered equation. Its consequences were not deduced in that reference.

\refstepcounter{remark} \label{r.bg2}
\smallskip\noindent\textsf{Remark \arabic{remark}}:~
Note that it is the initial temperature that appears in \Eqsref{e.bg3.13}{e.bg3.14}\@. Moreover, $\rho_A\U$ is not in general an equilibrium state.

\smallskip

Suppose that $B$ starts in an arbitrary initial state $\rho_{B,0}$, while $A$ begins in the thermal state $\rho_{A,\beta_A}$\@. Combining Eqs.\ (\ref{e.bg3.9}) and (\ref{e.bg3.13}) and assuming that $U$ preserves entropy, we obtain
\begin{equation}
	\beta_A \du E_A(\rho) + \du S(\rho_B) =
	S(\rho\U|\rho_A\U\otimes \rho_B\U) +
	S(\rho_A\U|\rho_{A,\beta_A})
\,.
\label{e.bg3.15}
\end{equation}
The last equation requires that $U$ preserve entropy. It also remains valid if the Hamiltonian of $B$, $H_B$, depends on an external parameter varying with time, so that $B$ receives work from an external agent. On the other hand, $H_A$ should be time independent. Then if $U$ conserves energy
\begin{equation}
	\beta_A \left( \du E_B(\rho) + \du E_V(\rho) \right) = \du
	S(\rho_B) - \left[ S(\rho\U|\rho_A\U\otimes \rho_B\U) +
	S(\rho_A\U|\rho_{A,\beta_A}) \right].
\label{e.bg3.16}
\end{equation}
These relations imply the following inequalities:\\
1) If $U$ preserves entropy, even if $H_B$ depends on an external parameter varying with time
\be
	\beta_A \du E_A(\rho) \ge - \du S(\rho_B)
\label{e.bg3.17}
\,.
\ee
2) If $U$ conserves entropy and the total energy, one has
\be
	\beta_A\left( \du E_B(\rho) + \du E_V(\rho) \right) \le \du
	S(\rho_B)
\label{e.bg3.18}
\,,
\ee
with the following interpretations. Suppose $U$ conserves the entropy; then we couple a system $B$ (initially in an arbitrary state $\rho_{B,0}$) to system $A$ (initially in thermal equilibrium) and that we want to lower the entropy of $B$ so that $\du S(\rho_B) \le 0$\@. Then, the energy of $A$ must increase by at least
\begin{equation}
	\du E_A(\rho) \ge \frac{1}{\beta_A}|\du S(\rho_B)|
\label{e.bg3.19}
\end{equation}
even if $B$ receives work from an external source (so that $H_B$ depends on an external parameter). Moreover, if the total energy is conserved, the sum of the energy of $B$ and the coupling energy must decrease by at least:
\begin{equation}
	\du E_B(\rho) + \du E_V(\rho) \le \frac{1}{\beta_A}\du
	S(\rho_B) < 0
\,.
\label{e.bg3.20}
\end{equation}
Thus lowering the entropy of a system $B$, coupled to a system initially at equilibrium, costs transfers of energy from $B$ to $A$ or to the interaction energy, a result analogous to those of Brillouin \cite{brillouin2} and Landauer \cite{landauerIBM1961}, even if system $B$ receives work from an external source. But note again that only the temperature $\beta_A$ appears. This is the initial temperature at the beginning of the evolution $U$, so that system $A$ is not necessarily a thermal bath, because its temperature may vary during the evolution~$U$\@.

\subsection{The case of equality in \Eqref{e.bg3.19}\label{s.bg3.4}}

Suppose that $\delta\U S(\rho_B)<0$ and that one has equality in \Eqref{e.bg3.19} or \Eqref{e.bg3.17}. Then by \Eqref{e.bg3.15} one has $S(\rho\U|\rho\U_A \otimes \rho\U_B) = S(\rho\U_A|\rho_{A,\beta_A})=0$\@.
This implies $\rho\U_A=\rho_{A,\beta_A}$ $\rho\U=\rho\U_A\otimes \rho\U_B=\rho_{A\beta_A} \otimes \rho\U_B$ then $\delta\U S(\rho_B)=0$ and $\delta E_A(\rho_A)=0$, and from \Eqref{e.bg3.9}, and therefore the entropy of $B$ has not changed. Thus lowering the entropy of $B$ is inconsistent with equalities in \Eqref{e.bg3.19}.

\subsection{Both systems $A$ and $B$, are at equilibrium \label{s.bg3.5}}

Assume that $A$ and $B$ are initially at thermal equilibrium at different temperatures. Then, one has\\
1. For a general evolution
\be
\du S(\rho_A)-\beta_A \du E_A(\rho_A) = -S(\rho_A\U|\rho_{A,\beta_A})
\,,
\label{e.bg3.21}
\ee
\be
\du S(\rho_B)-\beta_B \du E_B(\rho_B) = -S(\rho_B\U|\rho_{B,\beta_B})
\,.
\label{e.bg3.22}
\ee
2. If $U$ conserves entropy
\be
\du S(\rho_A)+\du S(\rho_A) = S(\rho\U|\rho_{A}\U\otimes\rho_{B}\U)
\,.
\label{e.bg3.23}
\ee
3. If $U$ conserves energy
\be
\du E_A(\rho_A)+\du E_B(\rho_B)+\du E_V(\rho) = 0
\,.
\label{e.bg3.24}
\ee
Then, we conclude\\
A. If $U$ conserves entropy: Combining Eqs.\ (\ref{e.bg3.21}), (\ref{e.bg3.22}), and (\ref{e.bg3.23}) yields
\be
\beta_A\du E_A(\rho_A)+\beta_B\du E_B(\rho_B) = S(\rho\U|\rho_{A,\beta_A}\otimes\rho_{B,\beta_B})
\,,
\label{e.bg3.25}
\ee
\bea
S(\rho\U|\rho_{A,\beta_A}\otimes\rho_{B,\beta_B})&=& \du S(\rho_A)+\du S(\rho_B)   \qqquad\qqqqquad~
  \nonumber\\
  & =&\beta_A \du S(\rho_A) +\beta_B \du S(\rho_B) -\left[S(\rho_A\U|\rho_{A,\beta_A})+S(\rho_B\U|\rho_{B,\beta_B})
                          \right]
.
\label{e.bg3.26}
\eea
This last identity implies the Clausius inequality
\be
0\le \du S(\rho_A)+\du S(\rho_B)
   \le \beta_A \du S(\rho_A) +\beta_B \du S(\rho_B)
\,.
\label{e.bg3.27}
\ee
B. For a general evolution $U$: Combining \Eqsref{e.bg3.21}{e.bg3.22}
\be
\du E_A(\rho_A)+\du E_B(\rho_B)=T_A \left[\du S(\rho_A)+S(\rho_A\U|\rho_{A,\beta_A})\right]
                     +T_B \left[\du S(\rho_B)+S(\rho_B\U|\rho_{B,\beta_B})\right]
\,,
\label{e.bg3.28}
\ee
and thus
\be
\du E_A(\rho_A)+\du E_B(\rho_B)\ge T_A \du S(\rho_A) +T_B \du S(\rho_B)
\,.
\label{e.bg3.29}
\ee

\subsection{Case of equality in \Eqsref{e.bg3.27}{e.bg3.29} \label{s.bg3.6}}

\noindent
A. $U$ conserves entropy.
Equality in \Eqref{e.bg3.27} implies immediately that $\rho\U_A=\rho_{A,\beta_A}$ and $\rho\U_B=\rho_{B,\beta_B}$,
in which case the energy of $A$ and the energy of $B$ have not changed and $\du S(\rho_A)=\du S(\rho_B)=0$\@. From the first equality in \Eqref{e.bg3.26} one has
\be
\rho\U = \rho_A\U\otimes\rho_B\U=\rho_{A,\beta_A}\otimes\rho_{B,\beta_B}
\,,
\label{e.bg3.30}
\ee
and one deduces that the state $\rho$ has not changed.

\noindent
B. General evolution $U$\@. If one has equality in \Eqref{e.bg3.29}, it follows from \Eqref{e.bg3.28} the same results as above: the state $\rho$ has not changed.

\subsection{Interaction energy and relative entropy \label{s.bg3.7}}

We assume that $U$ conserves entropy and energy. Divide \Eqsref{e.bg3.21}{e.bg3.22} by $\beta_B$ and add; then use the conservation of energy \Eqref{e.bg3.24} to eliminate $\delta\U E_B(\rho)$ and deduce after some calculations
\be
-\du E_V(\rho)=\left(1-\frac{\beta_A}{\beta_B}\right)\du E_A(\rho_A)+T_B S(\rho\U|\rho_{A,\beta_A}\otimes\rho_{B,\beta_B})
\,,
\label{e.bg3.31}
\ee
so that
\be
-\du E_V(\rho)\ge \left(1-\frac{\beta_A}{\beta_B}\right)\du E_A(\rho_A)
\,.
\label{e.bg3.32}
\ee
In case of equality in \Eqref{e.bg3.32}, one deduces that $\rho\U=\rho_{A,\beta_A} \otimes \rho_{B,\beta_B}$ so that the state has not changed and $\delta\U E_A(\rho_A)=\delta\U E_V(\rho)=0$\@. Moreover if $\delta\U E_A(\rho_A)$ is positive and $T_A$ is larger than  $T_B$, the interaction energy $V$ is not necessarily zero and $\delta\U E_V(\rho)$ is negative.

Finally, if one could neglect the interaction energy, \Eqref{e.bg3.32} implies that energy flows from the hot to the cold system.

\subsection{The case $\beta_A=\beta_B$ \label{s.bg3.8}}

Again assume that $U$ conserves both entropy and energy. From \Eqref{e.bg3.31} and the conservation of energy, one deduces
\be
-\du E_V(\rho)=\du E_A(\rho_A)+\du E_B(\rho_B)=\frac1{\beta} S(\rho\U|\rho_{A,\beta_A}\otimes\rho_{B,\beta_B})
\,,
\label{e.bg3.33}
\ee
so that $\delta\U E_V(\rho)\le0$\@. Thus when $A$ and $B$ are initially at thermal equilibrium at the same temperature, the sum of the energies of $A$ and $B$ can only increase at the expense of the interaction energy~\cite{ratchet}.

\section{Two systems in interaction with a work source\label{s.bg4}}

\subsection{Hypotheses\label{s.bg4.1}}

We consider two systems $A$ and $B$ in interaction, with system $A$ coupled to a work source. We represent the action of the work source by parameters, collectively denoted by $\lambda$, so that $H_A=H_A(\lambda)$\@. Thus we assume that $H_B$ and $V$ are independent of $\lambda$\@. The action of the work source is given by an evolution of the parameters $\lambda(t)$ imposed by an external agent. The total system \AplusB\ has a unitary or symplectic evolution $U(t)$ depending explicitly on time-$t$\@. Clearly, $U(t)$ conserves entropy but does not conserve energy, and instead one has the identity
\begin{equation}
	\du E_A(\rho) + \du E_B(\rho) + \du E_V(\rho) + \du W = 0
\,,
\label{e.bg4.1}
\end{equation}
with the following notation
\begin{eqnarray}
	\du E_B(\rho) &=&  \tr\left( (\rho\U - \rho_0) H_B \right)
\label{e.bg4.2} \\
	\du E_V(\rho) &=&  \tr\left( (\rho\U - \rho_0) H_V \right)
\label{e.bg4.3}\\
	\du E_A(\rho) &=&  \tr\left( \rho\U H_A(\lambda\U) - \rho_0
	H_A(\lambda_0) \right)
\,.
\label{e.bg4.4}
\end{eqnarray}
Here, $\lambda_0$ is the initial value of the parameter $\lambda$ and $\lambda\U$ is its final value at the end of the evolution $U$, this being an abbreviation for $U(t)$, $t$ being the final time. Note that \Eqref{e.bg4.4} extends the definition given near \Eqref{e.bg2.11}. Such an extension is needed because we now allow changes in the Hamiltonian, represented by the additional variable $\lambda$\@. \Eqref{e.bg4.1} defines the work $\delta\U W$, which is taken to be positive if the source receives work from the system \AplusB\@.

We assume that the initial state at time-0 is
\begin{equation}
	\rho_0 = \rho_{A,\beta_A,\lambda_0} \otimes \rho_{B,\beta_B}
\,,
\label{e.bg4.5}
\end{equation}
with
\begin{equation}
	\rho_{A,\beta_A,\lambda_0} = \frac{e^{-\beta_A
	H_A(\lambda_0)}}{Z_A(\beta_A,\lambda_0)}
\,,
\label{e.bg4.6}
\end{equation}
\begin{equation}
	Z_A(\beta_A,\lambda_0) = \tr\left( e^{-\beta_A H_A(\lambda_0)} \right)
\,,
\label{e.bg4.7}
\end{equation}
and
\begin{equation}
	Z_A(\beta_A,\lambda_0) = \exp\left( -\beta_A
	F_A(\beta_A,\lambda_0) \right)
\,.
\label{e.bg4.8}
\end{equation}
Here, $F_A(\beta_A,\lambda_0)$ denotes the equilibrium free energy for A\@. For a general state $\rho$ of a system with energy $H$ we define the non equilibrium free energy of the state $\rho$ at temperature $T$ to be
\be
F(\beta,\rho)= \Tr(\rho H)-T S(\rho)
\,.
\label{e.bg4.8a}
\ee
In particular, for subsystem $A$ one can define the non equilibrium free energy of the state $\rho\U_A$ at temperature $T_A$ to be
\begin{equation}
	F_A\U(\beta_A) = \tr\left( \rho_A\U H_A(\lambda\U) \right) -
	\frac{1}{\beta_A}S(\rho_A\U)
\,.
\label{e.bg4.8b}
\end{equation}
In both of the above formulas temperature is not necessarily related to the state~$\rho$\@.

\subsection{Identities for the work\label{s.bg4.2}}

We next establish the following two relations
\bea
	\du W &=& -\du E_V(\rho) - \tr\left(\rho_A\U\left[H_A(\lambda\U)-H_A(\lambda_0)\right] \right)
            \nonumber\\ &&
        -\left(1-\frac{\beta_B}{\beta_A}\right) \du E_B(\rho) -
         \frac{1}{\beta_A}S(\rho\U|\rho_{A,\beta_A}\otimes \rho_{B,\beta_B})
\label{e.bg4.9}
\eea
and
\bea
		\du W = -\du E_V(\rho) - \left( 1-\frac{\beta_B}{\beta_A}
		\right)\du E_B(\rho) + \left( F_A(\beta_A,\lambda_0) - F_A\U \right)
                \nonumber\\
		- \frac{1}{\beta_A}\left( S(\rho\U|\rho_A\U\otimes
		\rho_B\U)+ S(\rho_B\U |\rho_{B,\beta_B})\right)
\,,
\label{e.bg4.10}
\eea
with $F\U_A$ the non equilibrium free energy of $\rho\U_A$ calculated at the initial temperature $T_A$, namely
\Eqref{e.bg4.8b},  $F\U_A= \Tr_A(\rho\U_A H_A(\lambda\U) -T_A S(\rho\U_A)$\@. We will comment on these relations in Par.~\ref{s.bg4.3}

\refstepcounter{remark} \label{r.bg.near4.10}
\smallskip\noindent\textsf{Remark \arabic{remark}}:~
Here the free energy of \Eqref{e.bg4.8b} is not a state function, because it is calculated at the initial temperature of A\@.

\smallskip
\noindent
Proof of \Eqref{e.bg4.9}: One again starts from the fundamental identities \Eqsref{e.bg2.6}{e.bg3.13}
\begin{equation}
	\du S(\rho_B) - \beta_B \du E_B(\rho) = -
	S(\rho_B\U|\rho_{B,\beta_B})
\,,
\label{e.bg4.12}
\end{equation}
\begin{equation}
	\du S(\rho_A) - \beta_A \tr_A\left( (\rho_A\U -
	\rho_{A,\beta_A,\lambda_0}) H_A(\lambda_0) \right) = -
	S(\rho_A\U|\rho_{A,\beta_A,\lambda_0})
\,,
\label{e.bg4.13}
\end{equation}
and
\begin{equation}
	\du S(\rho_A) + \du S(\rho_B) = S(\rho\U|\rho_A\U \otimes
	\rho_B\U)
\,.
\label{e.bg4.14}
\end{equation}
Add \Eqsref{e.bg4.12}{e.bg4.13} and subtract \Eqref{e.bg4.14}, using the fact that
\begin{equation}
	S(\rho\U| \rho_A\U\otimes \rho_B\U) +
	S(\rho_A\U|\rho_{A,0}) + S(\rho_B\U|\rho_{B,0}) =
	S(\rho\U|\rho_{A,0}\otimes \rho_{B,0})
\,.
\label{e.bg4.15}
\end{equation}
We obtain
\begin{equation}
	-\beta_B \du E_B(\rho) - \beta_A \tr_B\left( (\rho_A\U -
	\rho_{A,\beta_A,\lambda_0})H_A(\lambda_0) \right) = -
	S(\rho\U|\rho_{A,0}\otimes \rho_{B,0})
\,.
\label{e.bg4.16}
\end{equation}
Conservation of energy \Eqref{e.bg4.1} gives
\bea
	\du W &=& - \du E_V(\rho) - \du E_B(\rho)
           - \tr_A\left(\rho_A\U\left( H_A(\lambda\U) - H_A(\lambda_0) \right)\right)
           \nonumber\\
          && - \tr_A\left( \left( \rho_A\U - \rho_{A,\beta_A,\lambda_0} \right)H_A(\lambda_0) \right)
\,.
\label{e.bg4.17}
\eea
We eliminate the second trace in the right hand side of \Eqref{e.bg4.17} using \Eqref{e.bg4.16}, multiply by $T_A$ to obtain \Eqref{e.bg4.9}.

\smallskip\noindent
Proof of \Eqref{e.bg4.10}: In \Eqref{e.bg4.9}, we replace the relative entropy term, using
\bea
		S\left(\rho_A\U  \big\vert  \rho_{A,\beta_A,\lambda}\right)
        & =& -S(\rho_A\U) + \beta_A \tr\left( \rho_A\U 	H_A(\lambda_0) \right) + \log Z_A(\beta_A,\lambda)
\nonumber\\
		&=& -S(\rho_A\U) + \beta_A \tr\left(\rho_A\U H_A(\lambda_0) \right) + \beta_A F_A(\beta_A,\lambda_0)
\,,
\label{e.bg4.17a}
\eea
and use the definition of $F\U_A$ of \Eqref{e.bg4.8b}
\begin{equation}
	-\frac{1}{\beta_A}S\left(\rho_A\U\big\vert\rho_{A,\beta_A,\lambda_0}\right)
	- \tr\left( \rho_A\U\left( H_A(\lambda\U) - H_A(\lambda_0)
	\right) \right) = F_A(\beta_A,\lambda_0) - F_A\U
\,.
\label{e.bg4.18}
\end{equation}

\subsection{Inequalities for the work\label{s.bg4.3}}

From the identities of \Eqsref{e.bg4.9}{e.bg4.10}, we deduce immediately corresponding inequalities
\begin{equation}
	\du W \le - \du E_V(\rho) - \tr_A\left( \rho_A\U\left(
	H_A(\lambda\U) - H_A(\lambda_0)
	\right) \right) - \left( 1-\frac{\beta_B}{\beta_A} \right)\du
	E_B(\rho)
\label{e.bg4.19}
\end{equation}
and
\begin{equation}
	\du W \le - \du E_V(\rho) - \left( 1-\frac{\beta_B}{\beta_A}
	\right) \du E_B(\rho) + \left( F_A(\beta_A,\lambda_0) - F_A\U \right)
\,.
\label{e.bg4.20}
\end{equation}
The interpretation of inequality (\ref{e.bg4.20}) is straightforward. If one can neglect the interaction energy, and if $T_A =T_B$, one gets an analogue of the familiar thermodynamic inequality giving an upper bound between of the work received by the work source and the variation of the free energy of A,
\begin{equation}
	\du W \le F_A(\beta_A,\lambda_0) - F_A\U
\,.
\label{e.bg4.20a}
\end{equation}

\medskip
\refstepcounter{remark} \label{r.mm}
\smallskip\noindent\textsf{Remark \arabic{remark}}:~
\Eqref{e.bg4.10} contains much more information than inequalities \Eqsref{e.bg4.20}{e.bg4.20a}, since it expresses the difference between the maximum work that can be delivered by system $A$ and the work effectively extracted from $A$, which is the energy dissipated in the process. It is expressed in terms of relative entropies, and it will be shown in Sec.\ \ref{s.bg6} that it can be explicitly estimated, which yields a calculation of transport coefficients from first principles.

\subsection{The case of equalities in \Eqsref{e.bg4.19}{e.bg4.20}\label{s.bg4.4}}
If one has equality in \Eqref{e.bg4.19}, the relative entropy of \Eqref{e.bg4.9} must be equal to 0,
\be
S\left(\rho\U  \big\vert\,\rho_{A,\beta_A}\otimes \rho_{B,\beta_B}\right) = 0
\,,
\label{e.bg4.21x}
\ee
so $\rho\U=\rho_{A,\beta_A}\otimes \rho_{B,\beta_B}$ and the final state has come back to its initial value. In general, this would be impossible except if the final value of the parameter $\lambda\U=\lambda_0$, in which case by \Eqref{e.bg4.9}, $\delta\U W=0$\@. If we have equality in \Eqref{e.bg4.20}, both relative entropies of \Eqref{e.bg4.10} are equal to 0\@. In this case $\rho\U_B$ has come back to its initial value $\rho_{B,\beta_B}$ and $\delta\U E_B(\rho)=0$\@. Then, one has
\begin{equation}
	\du W = - \du E_V(\rho) + F_A(\beta_A,\lambda_0) - F_A\U
\,.
\label{e.bg4.21}
\end{equation}

\subsection{Case where $A$ is not initially in thermal equilibrium.\label{s.bg4.5}}

We shall now assume that the initial state is
\begin{equation}
\rho_0=\rho_{A,0}\otimes\rho_{B,\beta_0}
\,,
\label{e.bg4.22}
\end{equation}
$\rho_{A,0}$ being a general state.

The following identity also holds:
\begin{equation}
	-\du W = \du F_A(\beta_B,\rho_A) +\du E_V(\rho) + T_B
	\left(S(\rho\U|\rho_A\U \otimes \rho_B\U) +
	S(\rho_B\U|\rho_{B,\beta_B})\right)
\,.
\label{e.bg4.23}
\end{equation}
This equation can be found in \cite{esposito1, esposito2}. It is true even if $A$ is not initially in a thermal state. Note the temperature of $B$ appearing in the non equilibrium free energy of~$A$
\begin{equation}
	F_A(\beta_B,\rho_A) = E_A(\rho_A) - T_B S(\rho_A)
\,.
\label{e.bg4.24}
\end{equation}
If no work is performed, \Eqref{e.bg4.23} reduces to \Eqref{e.bg3.16} upon exchanging the labels $A$ and $B$\@.

\smallskip
\noindent
Proof: Using $S(\rho\U)=S(\rho_0)$, one has
\begin{equation}
S(\rho\U|\rho_A\U\otimes\rho_B)+S(\rho_B\U|\rho_{B,\beta_B})=
-S(\rho_{A,0})-S(\rho_{B,\beta_B})+S(\rho_A\U)+\beta_B E_B(\rho_B\U) + \log Z_B(\beta_B)
\,.
\label{e.bg4.25}
\end{equation}
Then,
\begin{equation}
\log Z_B(\beta_B)= -\beta_B E_B(\rho_{B,\beta_B}) + S(\rho_{B,\beta_B})
\,,
\label{e.bg4.26}
\end{equation}
and \Eqref{e.bg4.25} becomes
\begin{equation}
S(\rho\U|\rho_A\U\otimes\rho_B)+S(\rho_B\U|\rho_{B,\beta_B})=
\du S(\rho_A)+\beta_B \du E_B(\rho_B)
\,.
\label{e.bg4.27}
\end{equation}
Using the conservation of energy, \Eqref{e.bg4.1}, one obtains
\begin{equation}
\du W=-\du E_V(\rho) -\du F_A(\beta_B,\rho_A) - T_B
           \left[ S(\rho\U|\rho_A\U\otimes\rho_B)+S(\rho_B\U|\rho_{B,\beta_B})
           \right]
\,.
\label{e.bg4.28}
\end{equation}
Here
\begin{equation}
\du F_A(\beta_B,\rho_A)=\du E_A(\rho_A) -T_B \du S(\rho_A)
\label{e.bg4.29}
\end{equation}
is the variation of the non equilibrium free energy of $A$ calculated at the initial temperature $T_B$ of $B$ and
\begin{equation}
\du E_A(\rho_A)=\Tr_A\left( H_A(\lambda\U)\rho_A\U \right)-\Tr_A\left( H_A(\lambda_0)\rho_{A,0} \right)
\,.
\label{e.bg4.30}
\end{equation}
In particular
\begin{equation}
\du W \le -\du E_V(\rho) -\du F_A(\beta_B,\rho_A)
\,,
\label{e.bg4.31}
\end{equation}
with equality if and only if the two relative entropy terms of \Eqref{e.bg4.28} are zero, which means that
\begin{equation}
\rho_B\U=\rho_{B,\beta_B} \quad \hbox{and}\quad \rho\U=\rho_A\U\otimes\rho_{B,\beta_B}
\,.
\label{e.bg4.32}
\end{equation}

\smallskip
\refstepcounter{remark} \label{r.bg.near4.32}
\smallskip\noindent\textsf{Remark \arabic{remark}}:~
In general during the evolution $U$, the state of $B$ does not remain thermal because $B$ is not necessarily a thermal bath and correlations develop between $A$ and~B\@.

\section{A system coupled only to an external work source\label{s.bg5}}

\subsection{Hypotheses\label{s.bg5.1}}

We consider a system coupled only to an external work source, so that the Hamiltonian of the system is $H(\lambda)$\@.

At time $t = 0$, the state of the system is supposed to be a thermal state $\rho_{\beta_0}(\lambda_0)$\@. The external observer imposes an evolution $\lambda(t)$ of the parameter $\lambda$ from $\lambda_0$ to $\lambda\U$, inducing a unitary or symplectic evolution $U$ of the whole system. The work that the external observer must perform to realize this evolution is obviously the variation of the energy of the system. With the convention of Sec.\ \ref{s.bg4.1}, we denote by $\du W$ the work counted positive if the system receives it from the external source. We are now in a particular case of Sec.\ \ref{s.bg4.1} when the system is $A$, there is no system $B$ and no $V$\@. Thus from \Eqref{e.bg4.1}
\begin{equation}
	\du W = - \du E(\rho) = \tr\biggl( \rho_{\beta_0}(\lambda_0)
	H(\lambda_0) - \rho\U H(\lambda\U) \biggr).
\label{e.bg5.1}
\end{equation}

\subsection{Identities for the work\label{s.bg5.2}}

From \Eqsref{e.bg4.9}{e.bg4.10} we obtain immediately
\begin{equation}
	\du W = - \tr\biggl( \rho\U \left( H(\lambda\U) - H(\lambda_0)
	\right) \biggr) - \frac{1}{\beta_0}
	S\left(\rho\U|\rho_{\beta_0}(\lambda_0)\right)
\label{e.bg5.2}
\end{equation}
and
\begin{equation}
	\du W = F(\beta_0,\lambda_0) - F\U
\,,
\label{e.bg5.3}
\end{equation}
with $F\U$ the non equilibrium free energy at temperature $\beta_0$\@.
\begin{equation}
	F\U  = \tr\left( \rho\U H(\lambda\U) \right) -
	\frac{1}{\beta_0} S(\rho\U)
\label{e.bg5.4}
\end{equation}
We now prove the following identity
\begin{equation}
	\du W = F(\beta_0,\lambda_0) - F(\beta_0,\lambda\U) -
	\frac{1}{\beta_0}S(\rho\U|\rho_{\beta_0}(\lambda\U))
\,.
\label{e.bg5.5}
\end{equation}
This is a particular case of the result of \cite{kawai}\@. \\

\medskip\noindent
Proof of \Eqref{e.bg5.5}: We start from equation \Eqref{e.bg5.3} written as
\begin{equation}
	\du W = F(\beta_0,\lambda_0) - F\U = F(\beta_0,\lambda_0) -
	F(\beta_0,\lambda\U) + F(\beta_0,\lambda\U) - F\U
\,.
\label{e.bg5.6}
\end{equation}
Now
\begin{equation}
	\beta_0 \left( F\U - F(\beta_0,\lambda\U) \right) = -
	S(\rho\U) + \beta_0 \tr\left( H(\lambda\U)\rho\U \right) -
	\beta_0 F(\beta_0,\lambda\U)
\,.
\label{e.bg5.7}
\end{equation}
But
\begin{equation}
	S(\rho\U|\rho_{\beta_0}(\lambda\U)) = - S(\rho\U) + \beta_0
	\tr\left( H(\lambda\U)\rho\U \right) + \log
	Z(\beta_0,\lambda\U)
\,,
\label{e.bg5.8}
\end{equation}
so that comparing Eqs.\ (\ref{e.bg5.7}) and (\ref{e.bg5.8}), one has
\begin{equation}
	\beta_0\left( F\U - F(\beta_0,\lambda\U) \right) =
	S(\rho\U|\rho_{\beta_0}(\lambda\U))
\,,
\label{e.bg5.9}
\end{equation}
and from \Eqref{e.bg5.6} we then deduce \Eqref{e.bg5.5}.

\smallskip
\refstepcounter{remark} \label{r.bg.near5.9}
\smallskip\noindent\textsf{Remark \arabic{remark}}:~
Since the transition under discussion is adiabatic, free energy (as in \Eqref{e.bg5.5}) is less suitable than internal energy. See Subsec.~\ref{s.bg5.6}\@.

\subsection{Inequalities for the work\label{s.bg5.3}}

\subsubsection{From \Eqref{e.bg5.2}\label{s.bg5.3.1}}

From \Eqref{e.bg5.2} we deduce
\begin{equation}
	\du W \le -\tr\left( \rho\U\left( H(\lambda\U) - H(\lambda_0) \right) \right)
\,,
\label{e.bg5.10}
\end{equation}
with equality if and only if $\rho\U = \rho_{\beta_0}(\lambda_0)$, i.e., the final state $\rho\U$ is the initial state.

\subsubsection{From \Eqref{e.bg5.5}\label{s.bg5.3.2}}

From \Eqref{e.bg5.5} we deduce
\begin{equation}
	\du W \le F(\beta_0,\lambda_0) - F(\beta_0,\lambda\U)
\label{e.bg5.11}
\end{equation}
with equality if and only if
\begin{equation}
	\rho\U = \rho_{\beta_0}(\lambda\U)
\,.
\label{e.bg5.12}
\end{equation}
That is, $\rho\U$ is the thermal state at the initial temperature and final value $\lambda\U$ of $\lambda$\@. Note that a necessary condition for this is that the entropy of the final thermal state is the same as the entropy of the initial state.

\subsection{Relation to the identity of Jarzynski\label{s.bg5.4}}

Let $z$ denote a point in the phase space of the system. In this section we assume that the dynamics is classical.

We denote by $z(s|z_0)$ the classical trajectory of the phase space point at time $s$ starting from $z_0$ at time $s=0$, for the classical evolution $U$\@. The external observer imposes the variation $\lambda(s)$ of $\lambda$ from $\lambda_0$ to $\lambda\U = \lambda(t)$\@. The identity of Jarzynski is \cite{jarzynski}:
\begin{equation}
	\left\langle e^{-\beta_0 \left( H(z(t|z_0),\lambda\U) -
	H(z_0,\lambda) \right)}\right\rangle_{\rho_{\beta_0}(\lambda_0)} =
	\frac{Z(\beta_0,\lambda\U)}{Z(\beta_0,\lambda_0)} = \exp\left(
	-\beta_0\left( F(\beta_0,\lambda\U) - F(\beta_0,\lambda_0) \right) \right)
\,.
\label{e.bg5.13}
\end{equation}
Because the exponential function is strictly convex, Jensen's inequality implies that
\begin{equation}
	\exp\left( -\beta_0\left\langle H(z(t|z_0),\lambda\U) -
	H(z_0,\lambda_0)\right\rangle_{\rho_{\beta_0}(\lambda_0)} \right)
	\le \left\langle e^{-\beta_0 \left( H(z(t|z_0),\lambda\U) -
	H(z_0,\lambda)
	\right)}\right\rangle_{\rho_{\beta_0}(\lambda_0)}
\,,
\label{e.bg5.14}
\end{equation}
so that using \Eqref{e.bg5.13} and taking the logarithm, one obtains
\begin{equation}
	\du W \le F(\beta_0,\lambda_0) - F(\beta_0,\lambda\U)
\,,
\label{e.bg5.15}
\end{equation}
which is the inequality (\ref{e.bg5.11}).

But if the inequality (\ref{e.bg5.15}) is an equality, we deduce $\rho\U = \rho_{\beta_0}(\lambda\U)$ as in \Eqref{e.bg5.12}, but we also deduce that the inequality of Jensen (\ref{e.bg5.14}) is an equality. Because the exponential function is strictly convex, this implies that the differences
\begin{equation}
	H(z(t|z_0),\lambda\U) - H(z_0,\lambda_0) = C
\,,
\label{e.bg5.16}
\end{equation}
where $C$ is a constant independent of $z_0$ (but obviously dependent on $\lambda_0$, $\lambda\U$ and $t$); in other words, the ``microscopic work'' is independent of the microscopic trajectory. Although this equality would seem impossible, it turns out that identity (\ref{e.bg5.16}) can be realized for certain systems and evolutions of $\lambda$ (see appendix~A).

\subsection{Effective temperatures\label{s.bg5.5}}

Let $H(\lambda)$ be a Hamiltonian depending on $\lambda$ and $\rho$ a state (classical or quantum) with energy $E(\rho) = \tr\left( \rho H(\lambda) \right)$\@. We can define two temperatures for $\rho$\@.

\smallskip
\noindent
(i) The (inverse) temperature $\beta_e(\rho,\lambda)$ is the temperature \cite{note:suppressinverse} such that
\begin{equation}
	E(\rho) = E(\beta_e,\lambda)
\,,
\label{e.bg5.17}
\end{equation}
with
$E(\beta_0,\lambda) = \tr(\rho_{\beta_e}(\lambda) H(\lambda))$\@. It is known that $\partial E(\beta,\lambda)/\partial \beta <0 $, so that \Eqref{e.bg5.17} defines $\beta_e$ unambiguously. The basic identity (\ref{e.bg2.4}) shows that
\begin{equation}
	S(\rho_{\beta_e}(\lambda)) - S(\rho) =
	S(\rho|\rho_{\beta_e}(\lambda))
\,,
\label{e.bg5.18}
\end{equation}
so that
\begin{equation}
	S(\rho_{\beta_e}(\lambda)) \ge S(\rho)
\,,
\label{e.bg5.19}
\end{equation}
which is the well known fact that $\rho_{\beta_e}(\lambda)$ maximizes the entropy among all states $\rho$ having a fixed energy. The quantity $\beta_e(\rho,\lambda)$ can be called the {\em effective temperature}.

\smallskip
\noindent
(ii) There is a second temperature $\beta_a(\rho,\lambda)$ such that
\begin{equation}
	S(\rho) = S(\beta_a,\lambda)
\,.
\label{e.bg5.20}
\end{equation}
We call this the \textit{adiabatic temperature}, and by the same arguments as given above it is well-defined. From \Eqref{e.bg5.19} and \Eqref{e.bg5.20}, one has
\begin{equation}
	S(\beta_e,\lambda) \ge S(\beta_a,\lambda)
\,.
\label{e.bg5.21}
\end{equation}
Because
\begin{equation}
	\left.\frac{\partial E}{\partial
	S(\beta,\lambda)}\right|_{\lambda\;\textrm{fixed}} =
	\frac{1}{\beta}
\label{e.bg5.21a}
\end{equation}
we deduce from \Eqref{e.bg5.21} that
\begin{equation}
	E(\beta_a,\lambda) \le E(\beta_e,\lambda) = E(\rho)
\label{e.bg5.22}
\end{equation}
and
\begin{equation}
	\beta_a \ge \beta_e
\,.
\label{e.bg5.23}
\end{equation}
Because $S$ is a strictly increasing function of $E$ (for $\lambda$ fixed), one sees that in equation \Eqref{e.bg5.21} or (\ref{e.bg5.22}), one has equality if and only if $\beta_a = \beta_e$\@. Moreover, one has the identity
\begin{equation}
	S(\rho|\rho_{\beta_a}(\lambda)) -
	S(\rho|\rho_{\beta_e}(\lambda)) =
	S(\rho_{\beta_e}(\lambda)|\rho_{\beta_a}(\lambda))
\,,
\label{e.bg5.24}
\end{equation}
which can immediately be verified.

\subsection{A more precise expression for the work\label{s.bg5.6}}

In thermodynamics, for an adiabatic evolution, the work is related to the internal energy by $dE= dW$ rather, than to the free energy, and to the adiabatic temperature, rather than to the effective energy temperature.

Given the state $\rho\U$ (corresponding to the evolution $U$, the parameter varying from $\lambda_0$ to $\lambda\U$) we can define the adiabatic temperature $\beta_a\U$ such that
\begin{equation}
	S(\beta_a\U,\lambda\U) = S(\rho\U) = S(\beta_0,\lambda_0)
\,.
\label{e.bg5.25}
\end{equation}
We prove the following identity
\begin{equation}
	\du W = E(\beta_0,\lambda_0) - E(\beta_a\U,\lambda\U) -
	\frac{1}{\beta_a\U}S(\rho\U|\rho_{\beta_a\U}(\lambda\U))
\,.
\label{e.bg5.26}
\end{equation}

\smallskip
\noindent
Proof of \Eqref{e.bg5.26}: One has by definition (\ref{e.bg5.1})
\begin{equation}
	-\du W = E(\rho\U) - E(\beta_0,\lambda_0) = E(\rho\U) -
	E(\beta_a\U,\lambda\U) + E(\beta_a\U,\lambda\U) -
	E(\beta_0,\lambda_0)
\,.
\label{e.bg5.27}
\end{equation}
Then
\bea
	S\left(\rho\U\big\vert\,\rho_{\beta_a\U}(\lambda\U)\right) &=& - S(\rho\U) +
	\beta_a\U E(\rho\U) + \log Z(\beta_a\U,\lambda\U)
\nonumber\\
    &=& \beta_a\U
	\left( E(\rho\U) - E(\beta_a\U,\lambda\U) \right)
\,,
\label{e.bg5.27a}
\eea
because $S(\rho\U) = S(\beta_a\U,\lambda\U)$ by the definition (\ref{e.bg5.25}). From this result and \Eqref{e.bg5.27} we deduce \Eqref{e.bg5.26}.

As a consequence of \Eqref{e.bg5.26}, we deduce the inequality
\begin{equation}
	\du W \le E(\beta_0,\lambda_0) - E(\beta_a\U,\lambda\U)
\,.
\label{e.bg5.28}
\end{equation}
In standard thermodynamics, for system thermally isolated and coupled to a work source, one has $\ud E = \ud W$, because $\du S=0$ for an adiabatic (thermally isolated) process and we recover equality in \Eqref{e.bg5.28}. In this situation, the inequality (\ref{e.bg5.11}) comparing the work to the difference of free energies is not relevant, because the temperature does not remain constant.

Note that the bound upper bound (\ref{e.bg5.28}), given in terms of energy and the adiabatic temperature, is sharper than the bound given by (\ref{e.bg5.11}), which is in terms of free energy. This is proved in the next subsection.

\subsection{Upper bounds on the work delivered by a system. Comparison of \Eqsref{e.bg5.11}{e.bg5.28}\label{s.mm}}

We next show that using internal energy for the work inequality gives a sharper result than using the free energy. Specifically,
\begin{equation}
	\du W \le E(\beta_0,\lambda_0) - E(\beta_a\U,\lambda\U) \le F(\beta_0,\lambda_0)-F(\beta_0,\lambda\U)
\,.
\label{e.bg5.28.01}
\end{equation}
Proof of \Eqref{e.bg5.28.01}: We need only prove that
\be \Delta\equiv F(\beta_0,\lambda_0) -E(\beta_0,\lambda_0) - \left(F(\beta_0,\lambda\U) -E(\beta_a\U,\lambda\U) \right) \ge0
\,.
\label{e.bg5.28.02}
\ee
Using the definition of the equilibrium free energy and \Eqref{e.bg5.25} we have
\be
\Delta=-\frac1{\beta_0}\left[ S(\beta_a\U,\lambda\U)-S(\beta_0,\lambda\U)
                      \right]
                     + \left[ E(\beta_a\U,\lambda\U)-E(\beta_0,\lambda\U)
                      \right]
\,.
\label{e.bg5.28.03}
\ee
Note that in \Eqref{e.bg5.28.03} \textit{all} terms involving $\lambda$ are evaluated at $\lambda\U$\@. Therefore
\be
\Delta=\int_{\beta_0}^{\beta_a\U}   \left[
                \frac{\partial E(\beta,\lambda\U)}{\partial\beta}
                -\frac1{\beta_0}\frac{S(\beta,\lambda\U)}{\partial\beta}
                                    \right]d\beta
\,.
\label{e.bg5.28.04}
\ee
But
\be
\frac{\partial S(\beta,\lambda)}{\partial\beta} =
\frac{\partial S}{\partial E}\;\frac{\partial E(\beta,\lambda)}{\partial\beta}
=\beta\frac{\partial E(\beta,\lambda)}{\partial\beta}
\,.
\label{e.bg5.28.05}
\ee
Using \Eqref{e.bg5.28.05} in \Eqref{e.bg5.28.04}, we obtain
\be
\Delta=\int_{\beta_0}^{\beta_a\U}
                \frac{\partial E(\beta,\lambda\U)}{\partial\beta}
                \left[1-\frac{\beta}{\beta_0}   \right]d\beta
\,.
\label{e.bg5.28.06}
\ee
But $\frac{\partial E(\beta,\lambda\U)}{\partial\beta}<0$, so that $\Delta\ge0$\@. Note that this does not depend on which of $\beta_a$ and $\beta_a\U$ is larger.

\subsection{The case of equalities in Eqs.\ (\ref{e.bg5.28}) and (\ref{e.bg5.11})\label{s.bg5.7}}

\subsubsection{Equality in \Eqref{e.bg5.28}\label{s.bg5.7a}}

In this case, one has $S(\rho\U|\rho_{\beta_a\U}(\lambda\U)) = 0$ in \Eqref{e.bg5.26} so
\begin{equation}
	\rho\U = \rho_{\beta_a\U}(\lambda\U)
\,.
\label{e.s5.7.01}
\end{equation}
In particular, $\rho\U$ is a thermal state so that
\begin{equation}
	\beta_e\U = \beta_a\U
\,.
\label{e.s5.7.02}
\end{equation}
However, if one has equality in \Eqref{e.bg5.28}, this does not improve the upper bound of \Eqref{e.bg5.11} for the free energy,
\begin{equation}
	\du W \le F(\beta_0,\lambda_0) - F(\beta_0,\lambda\U)
\,.
\label{e.s5.7.03}
\end{equation}

\subsubsection{Equality in \Eqref{e.bg5.11}\label{s.bg5.7b}}

From \Eqref{e.bg5.12} we deduce that
\begin{equation}
	\rho\U = \rho_{\beta_0}(\lambda\U)
\,,
\label{e.s5.7.04}
\end{equation}
so that $\rho\U$ is a thermal state and thus
\begin{equation}
	\beta_e\U = \beta_0 = \beta_a\U
\,.
\label{e.s5.7.05}
\end{equation}
This implies that we also have equality in \Eqref{e.bg5.28}
\begin{equation}
	\du W = E(\beta_0,\lambda_0) - E(\beta_0,\lambda\U)
\,.
\label{e.s5.7.06}
\end{equation}

\subsection{The case $\lambda\U = \lambda_0$\label{s.bg5.8}}

If one assumes that the final value $\lambda\U$ of $\lambda$ is equal to its initial value, we see immediately that $\beta_a\U = \beta_0$\@. Indeed
\begin{equation}
	S(\beta_a\U,\lambda_0) = S(\rho\U) = S(\beta_0,\lambda_0)
\,,
\label{e.bg5.29x}
\end{equation}
so that the temperatures are equal $\beta_a\U = \beta_0$\@. In this case, one has from \Eqref{e.bg5.26}
\begin{equation}
	\du W = -\frac{1}{\beta_0}S(\rho\U|\rho_{\beta_0}(\lambda_0))
	\le 0
\,,
\label{e.bg5.29}
\end{equation}
with equality if and only if
\begin{equation}
	\rho\U = \rho_{\beta_0}(\lambda_0)
\,,
\label{e.bg5.29a}
\end{equation}
so that the state has returned to its initial value.

\refstepcounter{remark} \label{r.bg3}
\smallskip\noindent\textsf{Remark \arabic{remark}}:~
If the external observer imposes a variation $\lambda(t)$ of the control parameter with $\lambda(0) = \lambda_0$, $\lambda(t_1) =\lambda_1$, $\lambda(t_{\rm f}) = \lambda_0$, inequality (\ref{e.bg5.29}) says that at the end of the cycle, the observer has always lost work. In particular, the work that the external observer has put in the system in the time interval $[0,t_1]$ cannot be entirely recovered in the time interval $[t_1,t_{\rm f}]$ whatever one does, except if the final state $\rho\U$ is the initial state.

\refstepcounter{remark} \label{r.bg4}
\smallskip\noindent\textsf{Remark \arabic{remark}}:~
When $\lambda\U = \lambda_0$, one can also recover \Eqref{e.bg5.29} from the identity (\ref{e.bg5.5}). This identity reduces to
\begin{equation}
	\du W = -\frac{1}{\beta_0}S(\rho\U|\rho_{\beta_0}(\lambda_0))
\,.
\label{e.bg5.29b}
\end{equation}

\section{Relative entropies and interaction, Fourier's law\label{s.bg6}}

In this Section we derive dissipation in the quantum context and show it to be intimately related to the relative entropy.

\subsection{The Born approximation\label{s.bg6.1}}

A quantum system has a Hamiltonian
\begin{equation}
	H = H_0 + V.
\label{e.bg6.1x}
\end{equation}
Let $\psi_k^{(0)}$, $E_k^{(0)}$ be the eigenstates and eigenvalues of $H_0$\@. In the Born approximation, the state $\kpno$ becomes at time $t$ a state $|\psi_n(t)\rangle$ with
\begin{equation}
	|\psi_n(t)\rangle = \sum_k \akn(t)\; e^{-i \eko t/\hbar}\kpko
\,.
\label{e.bg6.1}
\end{equation}
The quantities $\akn(t) = \delta_{k,n} + \takn(t)$ satisfy
\begin{equation}
	i\hbar \frac{\ud \takn}{\ud t} = \sum_l V_{k,l}(t) \left(
	\delta_{l,n} + \taln
	\right)
\,,
\label{e.bg6.1a}
\end{equation}
where
\begin{equation}
	V_{k,l}(t) = V_{k,l}\exp\left( \frac{i}{\hbar}\left(\eko - \elo\right)t \right)
\,.
\label{e.bg6.1b}
\end{equation}
We assume here $V_{n,n} = 0$ for all $n$\@. One readily deduces that in the Born approximation
\begin{equation}
	\akn(t) = \frac{V_{k,n}}{\eko - \eno}\left( 1 - e^{i(\eko -
	\eno)\frac{t}{\hbar}} \right) \quad (k\ne n)
\label{e.bg6.2}
\end{equation}
and by unitarity $\sum_k |\akn(t)|^2 = 1$, so to second order in $\takn$
\begin{equation}
	2 \re \tann(t) = -\sum_{k\ne n}|\takn(t)|^2
\,.
\label{e.bg6.3}
\end{equation}
Let $\rho_0$ be an initial state diagonal in the basis $\psi_n^{(0)}$
\begin{equation}
	\rho_0 = \sum p_{0,n}|\psi_n^{(0)}\rangle \langle \psi_n^{(0)}|
\,.
\label{e.bg6.3a}
\end{equation}
Then, at time $t$, the state becomes
\bea
	\rho\Ut &=& \rho_0
           + \left( \sum_n p_{0,n}\sum_{l\ne n}e^{-i(\eno -	\elo) \frac{t}{\hbar}} \talnc \kpno \bplo + {\rm c.c.} \right)
	       \nonumber\\
	&&  + \sum_n p_{0,n}\left( \tann(t) + \tannc(t)\right) \kpno \bpno
           \nonumber\\
	&& + \sum_n p_{0,n} \sum_{k,l \ne n} \takn(t)\taln(t)^*
	e^{-i(\eko - \elo)\frac{t}{\hbar}}\kpko \bplo + \dots
\,.
\label{e.bg6.4}
\eea
If $L$ is a Hermitian operator diagonal in the basis $\psi_n^{(0)}$ with eigenvalues $\lambda_n$, using \Eqref{e.bg6.4} one obtains in the Born approximation
\begin{equation}
	\tr \left( L(\rho\Ut - \rho_0) \right) =
	\frac{1}{2}\sum_{k\ne n} |\takn(t)|^2
	(\lambda_k-\lambda_n)(p_{0,n}-p_{0,k})
\,.
\label{e.bg6.5}
\end{equation}

\subsection{Two interacting systems\label{s.bg6.2}}

We consider two quantum systems $A$, $B$ with Hamiltonians $H_A$, $H_B$ respectively, interacting. Denote by $V  = V_{A,B}$ the interaction energy and
\begin{equation}
	H = H_A + H_B + V
\,.
\label{e.bg6.6}
\end{equation}
We call $\kak$, $\eako$ (resp.~$\kbl$, $\eblo$) the eigenstates and eigenvalues of $H_A$ (resp.~$H_B$), and we apply the Born approximation to $H$, with $H_0 = H_A + H_B$\@. The non perturbed Hamiltonian $H_0$ has eigenstates $\kak\kbl$ with eigenvalues $\eako + \eblo$\@.

We assume that at time $t=0$, the state of the system $A+B$ is $\rho_0 = \rho_A \otimes \rho_B$ with
\begin{eqnarray}
	\rho_A & = & \sum p_{A,k}\kak \bak\nonumber\\
	\rho_B & = & \sum p_{B,l}\kbl \bbl
\,,
\label{e.bg6.7}
\end{eqnarray}
so that they are diagonal in the eigenbasis of $H_A$ and $H_B$ and therefore commute with $H_A + H_B$\@. At time $t$, the initial state $\rho_0 = \rho_A \otimes \rho_B$ evolves to $\rho(t)$\@. Then
\begin{equation}
	S(\rho(t) | \rho_A \otimes \rho_B) = \tr\left( \rho(t) \log
	\rho(t)
	\right) - \tr\left( \rho(t) \log (\rho_A \otimes \rho_B) \right)
\,.
\label{e.bg6.7a}
\end{equation}
But $S(\rho(t)) = S(\rho_0)$ by unitarity of the evolution, so that
\begin{equation}
	S(\rho(t) | \rho_A \otimes \rho_B)
           =
      -\tr\left[ \left( \rho(t)
	      - \rho_A \otimes \rho_B\right)(\log \rho_A + \log \rho_B)
	       \right]
\,.
\label{e.bg6.8}
\end{equation}
This is of the form of \Eqref{e.bg6.5} with
\begin{equation}
	L = -\left( \log \rho_A + \log \rho_B \right)
\,.
\label{e.bg6.8a}
\end{equation}
$L$ has eigenvectors $\kak\kbl$ with eigenvalues $\log p_{A,k} + \log p_{B,l}\,$; $\rho_A \otimes \rho_B$ has the same eigenvectors with eigenvalues $p_{A,k} + p_{B,l}$\@. Applying \Eqref{e.bg6.5}, one obtains in the Born approximation
\begin{equation}
	S(\rho(t) | \rho_A \otimes \rho_B) = \frac{1}{2}
	\sum_{(k,l)\ne (n,m)} |\taklnm(t)|^2 (p_{A,n}p_{B,m} -
	p_{A,k}p_{B,l}) \log \frac{p_{A,n}p_{B,m}}{p_{A,k}p_{B,l}}
\,.
\label{e.bg6.9}
\end{equation}
Notice that the quantity in the right hand side is automatically non-negative. Here, we have
\begin{equation}
	|\taklnm(t)|^2 = |\vklnm|^2 \
      \frac
      {\sin^2\left( \frac{\eako + \eblo - \eano - \ebmo}{2\hbar}\;t	\right)}
      {\left( \frac{\eako+\eblo-\eano-\ebmo}{2} \right)^2}
\,,
\label{e.bg6.10}
\end{equation}
with
$\vklnm = \langle \psi_{A,n}^{(0)}\otimes\psi_{B,m}^{(0)}|V| \psi_{A,k}^{(0)}\otimes\psi_{B,l}^{(0)}\rangle$\@.

We also deduce from this result that in this approximation $S(\rho(t)|\rho_A\otimes\rho_B) = 0$ if and only if $V  = 0$ (recall that the diagonal elements of $V$ are 0).

\subsection{The case where both initial states are thermal\label{s.bg6.3}}

Assume that at time $t=0$, $\rho_A = \rho_{A,\beta_A}$ and $\rho_B = \rho_{B,\beta_B}$ are the thermal states of $A$ and $B$ respectively. From \Eqref{e.bg3.25} one has
\begin{equation}
	\beta_A \du E_A(\rho_A) + \beta_B \du E_B(\rho_B) =
	S(\rho(t)|\rho_{A,\beta_A}\otimes\rho_{B,\beta_B}).
\label{e.bg6.11}
\end{equation}
Moreover, from conservation of energy
\begin{equation}
	\du E_A(\rho_A) + \du E_B(\rho_B) + \du E_V(\rho) = 0
\,,
\label{e.bg6.12}
\end{equation}
so that eliminating $\du E_B(\rho_B)$, one obtains
\begin{equation}
	(\beta_A - \beta_B) \du E_A(\rho_A) = \beta_B \du E_V(\rho) +
	S(\rho(t)|\rho_{A,\beta_A}\otimes \rho_{B,\beta_B})
\,.
\label{e.bg6.13}
\end{equation}
We now estimate both terms on the right hand side of~\Eqref{e.bg6.13}.

\subsubsection{Estimate of the relative entropy\label{s.bg6.3.1}}

From \Eqref{e.bg6.9} we deduce
\bea
	S(\rho(t) | \rho_A \otimes \rho_B) &=&
        \frac{1}{2Z_A Z_B} \sum_{(k,l)\ne (n,m)} |\taklnm(t)|^2 e^{-\left(\beta_A\ean +
		\beta_B\ebm\right)}
\nonumber \\		&&
         \times \left( 1- e^{-\left(\beta_A \left(\eak-\ean\right) +
		\beta_B\left(\ebl-\ebm\right)\right)} \right)
\nonumber\\ &&
        \times\left(\beta_A\left(\eak-\ean\right)
        +\beta_B\left(\ebl-\ebm\right) \right)
\,.
\label{e.bg6.14}
\eea
Moreover, when $t\rightarrow \infty$, as in the usual Born approximation, \Eqref{e.bg6.10} shows that
\begin{equation}
	|\taklnm(t)|^2 \simeq \frac{2\pi}{\hbar}|\vklnm|^2
	\delta\left(\eako + \eblo - \eano - \ebmo\right)t
\,.
\label{e.bg6.15}
\end{equation}
Thus if $f_A$ and $f_B$ denote the density of states for $A$ and $B$, we obtain from equation \Eqref{e.bg6.14}
\bea
		S(\rho(t) | \rho_A \otimes \rho_B) &=& \frac{2\pi}{2\hbar Z_A
		Z_B} \left( \beta_A - \beta_B \right)
		\int \ud E_A \ud E_A' \ud E_B \ud E_B' f_A(E_A) f_A\left( E_A'
		\right)
		f_B(E_B) f_B(E_B')
\nonumber\\&&
		\times e^{-\left( \beta_A E_A' + \beta_B E_B'
		\right)} |V_{(E_A,E_B)}^{(E_A',E_B')}|^2 \delta\left( E_A +
		E_B - E_A' - E_B'
		\right)\left( E_A - E_A' \right)
\nonumber\\&&
		\times\left( 1-e^{-\left( \beta_A -
		\beta_B
		\right)\left( E_A - E_A' \right)} \right)t
\,.
\label{e.bg6.16}
\eea

\subsubsection{Estimate of the interaction energy\label{s.bg6.3.2}}

Because $\du V(\rho) = -\du E_A(\rho) - \du E_B(\rho)$, one has
\begin{equation}
	\du V_{A,B}(\rho) = \tr\left(
	-(H_A+H_B)(\rho(t)-\rho_{A,\beta_A}\otimes \rho_{B,\beta_A})
	\right)
\,.
\label{e.bg6.17x}
\end{equation}
This is of the form of \Eqref{e.bg6.5} with $L = -H_A - H_B$, and so
\bea
		\beta_B \du V_{A,B}(\rho) &=& -\frac{\beta_B}{2 Z_A Z_B}
		\sum_{(k,l) \ne (n,m)} |\taklnm(t)|^2 \left( \eak + \ebl -
		\ean - \ebm \right)
\nonumber\\ &&
		\times e^{-\left( \beta_A E_A' + \beta_B E_B'
		\right)} \left( 1 - e^{-\left( \beta_A(\eak - \ean) +
		\beta_B\left( \ebl - \ebm \right) \right)} \right)t
\,.
\label{e.bg6.17}
\eea
Up to a sign, this expression is formally identical to the expression \Eqref{e.bg6.9}, \textit{except} that the difference of energies $(\eak + \ebl - \ean - \ebm)$ replaces the quantity $\beta_A(\eak-\ean) + \beta_B(\ebl - \ebm)$\@. As a consequence $\eak + \ebl - \ean - \ebm$ partially cancels the denominator of $|\taklnm(t)|^2$ and one sees that $\beta_B\du E_V(\rho)$ is negligible when $t\rightarrow \infty$\@.

Then from Eqs.\ (\ref{e.bg6.13}) and (\ref{e.bg6.16}), one sees that
\begin{equation}
	\du E_A(\rho_A) \simeq \frac{S(\rho(t)|\rho_{A,\beta_A}
	\otimes \rho_{B,\beta_B})}{\beta_A-\beta_B} \simeq (\beta_A -
	\beta_B)K t
\,.
\label{e.bg6.18}
\end{equation}
In \Eqref{e.bg6.18} $K$ is the positive constant
\begin{equation}
	K = \frac{2\pi}{2\hbar Z_A Z_B} \int \ud E_A \ud E_A' \ud E_B \ud
	E_B' \, \varphi(E_A,E_A',E_B,E_B')
\,,
\label{e.bg6.19}
\end{equation}
with
\bea
\varphi &=& f_A(E_A) f_A(E_A') f_B(E_B) f_B(E_B') e^{-\left(
		\beta_A E_A' +\beta_B E_B' \right)}
		\left|V_{(E_A,E_B)}^{(E_A',E_B')}\right|^2 \delta\left( E_A + E_B - E_A'
		- E_B'\right)
\nonumber \\&&
		\times\frac{E_A - E_A'}{\beta_A - \beta_B}\left( 1-
		e^{-\left( \beta_A - \beta_B \right)\left( E_A - E_A' \right)}
		\right)
\,.
\label{e.bg6.20}
\eea
It is obvious that $\varphi \ge 0$\@. Note that $K$ does not vanish for $\beta_A$ close to $\beta_B$\@.

The expression (\ref{e.bg6.18}) is a form of Fourier's law for heat transport from $B$ to $A$, $(\beta_A-\beta_B)K$ being the rate of dissipation. In this case, one sees that the significance of the relative entropy $S(\rho(t)|\rho_{A,\beta_A} \otimes \rho_{B,\beta_B})$ is that of a transport coefficient, here the transport of energy from one part of a system to another part.

\section{Coarse grained states\label{s.bg7}}

\subsection{Definition\label{s.bg7.1}}

Let $\rho$ and $\rho'$ be two states of the same system (classical or quantum). We say that $\rho'$ is obtained from $\rho$ by a coarse graining operation if
\begin{equation}
	\tr\left( \left( \rho - \rho' \right)\log \rho' \right) = 0.
\label{e.bg7.1}
\end{equation}
The idea is that the information associated with $\rho'$ (namely $\log\rho'$) is the same whether one averages with $\rho'$ or with the more detailed distribution~$\rho$\@. Using the basic identity, \Eqref{e.bg2.4}, we can say that $\rho'$ is obtained from $\rho$ by a coarse graining operation, if and only if
\begin{equation}
	S(\rho') - S(\rho) = S(\rho|\rho')
\,.
\label{e.bg7.2}
\end{equation}
In particular, $S(\rho')\ge S(\rho)$, so that the entropy increases by coarse-graining. (See the comment after \Eqref{e.bg2.3}.)

A coarse-graining mapping is a mapping $\Gamma$ which associates to any state $\rho$ (or to some states of a given class), a coarse grained state $\rho' = \Gamma(\rho)$\@.

\subsection{Examples of coarse-graining mappings\label{s.bg7.2}}

\noindent
\textsf{Example 1}: Maximum entropy.\\
Let $A_1,\dots,A_n$ be observables of the system, so they are either functions in the phase space or hermitian operators on the Hilbert space of the system. We consider the class of states $\rho$ such that
\begin{equation}
	\tr\left( A_i\rho \right) < \infty \,,\  \quad i = 1,..,n
\,.
\label{e.bg7.3x}
\end{equation}
One can then consider the state $\rho'$ such that $\rho'$ has maximal entropy given the relation
\be
\Tr\left(A_i\rho'\right)=\Tr\left(A_i\rho\right)
\,.
\label{e.bg7.3}
\ee
It is immediately seen that
\begin{equation}
	\rho' = C \exp\left( \sum_{i=1}^n \alpha_i A_i \right)
\,,
\label{e.bg7.4}
\end{equation}
where $C$ is a normalization constant and $\alpha_i$ are the ``conjugate parameters'', (provided $\rho'$ is normalizable). The mapping $\Gamma:\rho\rightarrow \rho'$ is indeed a coarse grain mapping in the sense of the previous definition, because by \Eqref{e.bg7.4}
\begin{equation}
	\tr\left( (\rho-\rho') \log \rho' \right) = \sum_{i=1}^n
	\tr\left( (\rho-\rho')A_i \right) = 0
\,.
\label{e.bg7.4a}
\end{equation}
In particular, one has \Eqref{e.bg7.2}.

The case of the thermal state is the best known, where one takes $A=H$, the Hamiltonian of the system.

\medskip
\noindent
\textsf{Example 2}: Naive coarse graining; the observables as characteristic functions.\\
(i) Classical case: Let $Z$ be the phase space of the system and $\{Z_i\}$ a finite partition of~$Z$ ($Z = \bigcup_{i=1}^n Z_i$ and $Z_i\bigcap Z_j=\emptyset$ for $i\ne j$). We choose $A_i = \chi_{Z_i}$ (i.e.~the characteristic function of $Z_i$). This is a particular case of example 1 and if $\rho$ is a state
\begin{equation}
	\rho' = C\exp\left( \sum_{i=1}^n \alpha_i \chi_{Z_i} \right)
\,.
\label{e.bg7.5}
\end{equation}
Using the condition (\ref{e.bg7.2}), namely,
\begin{equation}
	\int_{Z_i} \rho' \ud z = \int_{Z_i}\rho\,\ud z
\,,
\label{e.bg7.6}
\end{equation}
one can deduce from \Eqsref{e.bg7.5}{e.bg7.6}
\begin{equation}
	\rho'|_{Z_i} = \frac{1}{\textrm{Vol}(Z_i)} \int_{Z_i} \rho\,\ud z
\quad 	\textrm{or} \quad
\rho' = \sum_{i=1}^n
	(\rho'|_{Z_i})\chi_{Z_i}
\,.
\label{e.bg7.7}
\end{equation}
This equation implies that $\rho'$ is normalized $\int \rho' \ud z =1$\@. We recover the usual coarse graining.

\smallskip
\noindent
(ii) Quantum case: Let $\mathcal H$ be the Hilbert space of the system and $P_i$ a resolution of the identity by orthogonal projectors
\begin{equation}
	{\rm Id} = \sum P_i \quad\hbox{and}\quad P_i P_j = P_i\delta_{i,j}
\,.
\label{e.bg7.8x}
\end{equation}
Then the analogue of \Eqref{e.bg7.7} is
\begin{equation}
	\rho' = \sum_{i=1}^n \frac{\tr\left( \rho P_i \right)}{ {
	\rm dim} P_i({\mathcal H})}P_i
\,.
\label{e.bg7.8}
\end{equation}

\medskip
\noindent
\textsf{Example 3}: Coarse graining by marginals.\\
(i) Classical case: We assume that the system consists of several parts, and that its phase space is a Cartesian product, $Z = \prod_{i=1}^n Z_i$, corresponding to various subsystems with phase space $Z_i$\@. If $\rho$ is a state on $Z$, we denote by $\rho_i$ its marginal probability distribution on $Z_i$, so
\begin{equation}
	\rho_i(z_i) = \int\dots\int
	\rho(\zeta_1,\dots,\zeta_{i-1},z_i,\zeta_{i+1},\dots,\zeta_n)
	\prod_{j\ne i} \ud \zeta_j
\,.
\label{e.bg7.9}
\end{equation}
Let $\Gamma$ be the mapping that associates the product of its marginals to $\rho(z)$
\begin{equation}
	\Gamma(\rho)(z_1,\dots,z_n) = \prod_{i=1}^n \rho_i(z_i)
\,.
\label{e.bg7.10}
\end{equation}
Then the condition (\ref{e.bg7.1}) is satisfied. It is easy to see that $\prod_{i=1}^n \rho_i(z_i)$ is the state $\rho'$ that maximizes the entropy among all the states $\rho''$ such that $\rho_i'' = \rho_i$ for any $i$\@.

\medskip\noindent
(ii) Quantum case. The Hilbert space of the system is
\begin{equation}
	\mathcal{H} = \bigotimes_{i=1}^n \mathcal{H}_i
\,,
\label{e.A7.01}
\end{equation}
where the $\mathcal{H}_i$ are the Hilbert spaces of the subsystems. If $\rho$ is a state, then its marginal state on $\mathcal{H}_i$ is the partial trace on the Hilbert space $\mathcal{K}_i$, which is the tensor product of the Hilbert spaces $\mathcal{H}_j$ for $j$ different from~$i$
\begin{equation}
	\rho_i = \tr_{\mathcal{K}_i} \rho
\label{e.A7.02}
\end{equation}
and the mapping $\Gamma$,
\begin{equation}
	\Gamma(\rho) = \bigotimes_{i=1}^n \rho_i
\,,
\label{e.A7.03}
\end{equation}
is a coarse grained mapping. $\Gamma(\rho)$ is again the state $\rho'$ which maximizes the entropy among all states $\rho''$ such that $\rho_i'' = \rho_i$ for all $i$\@.

\medskip
\noindent
\textsf{Example 4}: Decomposition of $Z$\@. \\
If $Z = \bigcup_{i=1}^n Z_i$, but the $Z_i$ do not form a partition of $Z$ (they can have intersections of non-zero measure), one can still apply Example~1 to $A_i = \chi_{_{Z_i}}$ and obtain
\begin{equation}
	\rho' = C \exp\left( \sum \alpha_i \chi_{_{Z_i}} \right)
\,.
\label{e.A7.04}
\end{equation}
But now \Eqref{e.bg7.7} is no longer valid because, for given $z$, there will be in general several $i$ with $z\in Z_i$\@.

\subsection{Coarse graining and relative entropy\label{s.bg7.3}}

\noindent
(i) The case of the naive coarse-graining is distinguished among all types of coarse-graining by the following property. Let $Z= \bigcup_{i=1}^n Z_i$ a partition of the phase space and $p$, $q$ two probability distributions on $Z$\@. Let $\bar p = \Gamma p$ and $\bar q = \Gamma q$ be the coarse grained states of $p$ and $q$ associated to this partition. Then one has
\begin{equation}
	S(\bar p|\bar q) \le S(p|q).
\label{e.bg7.12}
\end{equation}
Proof: call $p_i = \int_{Z_i} p\ud z$ and $q_i = \int_{Z_i}q\ud z$\@. We have, using the definition of $\bar p$
and $\bar q$:
\begin{equation}
	S(\bar p|\bar q) = \sum_{i=1}^n p_i
	\log\frac{p_i}{q_i}
\,.
\label{e.bg7.13}
\end{equation}
Now
\bea
\sum_i p_i \log \frac{p_i}{q_i} &=& \sum_i
   \left( \int_{Z_i}p(z) \ud z \right)
        \log \frac{\int_{Z_i} p(z) \ud z}{\int_{Z_i}q(z')\ud
		z'}
\nonumber\\
		&=& \sum_i \left( \int_{Z_i}q(z') \ud z' \right)
          \left(
		\int_{Z_i} \frac{p(z)}{q(z)}\;\frac{q(z)}{\int_{Z_i}q(z')\ud z'}\ud
		z\right)
\nonumber\\ &&\quad\times
         \log \left( \int_{Z_i}
		\frac{p(z)}{q(z)}\;\frac{q(z)}{\int_{Z_i}q(z') \ud
		z'}\ud z \right)
\,.
\label{e.bg7.14}
\eea
But $\int_{Z_i} \frac{q(z)}{\int_{Z_i}q(z') \ud z'}\ud z = 1$\@. We use the fact that the function $x\log x$ is convex, so that for each $i$
\bea
&&	\left(
	\int_{Z_i}
	\frac{p(z)}{q(z)}\;\frac{q(z)}{\int_{Z_i}q(z')\ud
	z'}\ud
	z\right) \log
	\left(
	\int_{Z_i}
	\frac{p(z)}{q(z)}\;\frac{q(z)}{\int_{Z_i}q(z')
	\ud
	z'}\ud
	z
	\right)
\nonumber\\ &&\qquad\qquad\qquad
	\le
	\int_{Z_i} \frac{q(z)}{\int_{Z_i}q(z')\ud z'}\left(
	\frac{p(z)}{q(z)}\log \frac{p(z)}{q(z)} \;
	\right)\ud z
\,.
\label{e.bg7.15}
\eea
Therefore from \Eqref{e.bg7.15}
\be
\sum_i p_i \log\frac{p_i}{q_i} \le \sum_i \int_{Z_i} p(z) \log \frac{p(z)}{q(z)}\ud z = S(p|q)
\label{e.bg7.15a}
\,.
\ee

\medskip\noindent
(ii) For the coarse-graining associated to subsystems one has $Z = \prod_{i=1}^n Z_i$ and if $p$, $q$ are states on $Z$, the coarse grained states are $\tilde{p}=\bigotimes_{i=1}^n p_i$, $\tilde{q}=\bigotimes_{i=1}^n q_i$ and we deduce immediately that
\begin{equation}
	S(\tilde p|\tilde q) = \sum_{i=1}^n S(p_i|q_i)
\,.
\label{e.bg7.16}
\end{equation}

Consider the case $i=1$, and call $z = (z_1,z')$ with $z' = (z_2,\dots,z_n)$ and call
$Z' = Z_2 \times \dots\times Z_n$\@. Then
\bea
	S(p_1|q_1) &=& \int_{Z_1} \ud z_1
    \left(\int_{Z'} p(z_1,z') \ud z' \right) \log \frac{\int_{Z'} p(z_1,z')\ud
		z'}{\int_{Z'}q(z_1,z')\ud z'}
\nonumber \\ &=&
		\int_{Z_1}\ud z_1 \left( \int_{Z'}q(z_1,z'')\ud z''
		\right) \left( \int_{Z'}\ud z'
		\frac{p(z_1,z')}{q(z_1,z')}\;\frac{q(z_1,z')}{\int_{Z'}q(z_1,z'')\ud
		z''} \right)
\nonumber\\ && \qquad\qquad\qquad\times
\log \int_{Z'}\ud z'
		\frac{p(z_1,z')}{q(z_1,z')}\;\frac{q(z_1,z')}{\int_{Z'}q(z_1,z'')\ud
		z''}
\,.
\label{e.bg7.16a}
\eea
Now $\int_{Z'}\ud z' \frac{q(z_1,z')}{\int_{Z'}q(z_1,z'')\ud z''} = 1$\@. As in \Eqref{e.bg7.15}, we use the convexity of $x\log x$ and deduce that
\begin{equation}
	S(p_1|q_1) \le \int_{Z_1}\ud z_1 \int_{Z'}\ud z' p(z_1,z')
	\log \frac{p(z_1,z')}{q(z_1,z')} = S(p|q)
\,.
\label{e.bg7.16b}
\end{equation}

From \Eqref{e.bg7.16} we deduce that for the coarse graining mapping associated to the division of $Z = \prod_{i=1}^n Z_i$ in $n$ subsystems, one has
\begin{equation}
	S(\tilde{p}|\tilde{q}) \le n S(p|q)
\,.
\label{e.bg7.17}
\end{equation}

\medskip

\refstepcounter{remark} \label{r.bg5}
\smallskip\noindent\textsf{Remark \arabic{remark}}:~
The upper bound of \Eqref{e.bg7.17} cannot be improved. Indeed consider the case where: $p(z_1,\dots,z_n) = p_1(z_1)\delta(z_1-z_2)\dots\delta(z_{n-1}-z_n)$ $q(z_1,\dots,z_n) = q_1(z_1)\delta(z_1-z_2)\dots\delta(z_{n-1}-z_n)$\@. Then $p_i = p_1$ and $q_i = q_1$, but $S(p|q) = S(p_1|q_1)$ and $S(\bar p|\bar q) = n S(p_1|q_1)$\@.

\medskip\noindent
(iii) Thermal coarse graining.\\
Let $Z$ be a phase space, and $p$ and $q$ two probability distributions on $Z$, $H(z)$ a function of $z\in Z$\@. Let $\widetilde p$ and $\widetilde q$ be the thermal coarse grained probability distributions of $p$ and $q$, respectively, with respect to $H$\@. So
\begin{equation}
\widetilde p(z) =\frac1{Z(\beta(p))}\exp\left(-\beta(p)H(z)\right)
\,.
\label{e.bg7.17a}
\end{equation}
where $\beta(p)$ is the effective temperature of $p$, i.e., $\langle H\rangle_p=\langle H\rangle_{\widetilde p}$\@.

Assuming that $p-q$ is small, an obvious bound, after straightforward calculations, is
\begin{equation}
S(\widetilde p|\widetilde q)\le \frac{\langle H^2\rangle_p}{\langle H^2\rangle_{\widetilde p}-\left(\langle H \rangle_p\right)^2}
S(p|q)
\,.
\label{e.bg7.17b}
\end{equation}

This bound is surely not optimal, because if $p$ and $q$ are already thermal states, $S(\widetilde p|\widetilde q)=S( p| q)$\@. Note though that even without the hypotheses on $p$ and $q$, $S(\widetilde p|\widetilde q)\le S(p|\widetilde q)$\@.

\section{Conclusions}

The results in this article are used to obtain upper bounds for entropy production or energy variation in various situations of thermodynamic interest, with many such results either new or sharper than similar known bounds. Furthermore, the energy dissipated in these processes is expressed in terms of relative entropies, which  not only gives a general microscopic interpretation of dissipation, but also, in relevant examples, leads to an explicit, first principles, evaluation of dissipation terms, analogous to the Fourier law.

Although relative entropy has made appearances in many contexts, especially with respect to information theory, our results on a generalized Fourier heat law relates it in a direct way to the notion of dissipation as understood in physics.

\section*{Acknowledgements}
We (BG and LS) are grateful to the visitor program of the Max Planck Institute for the Physics of Complex Systems, Dresden, for its hospitality during the time that much of this work was performed.

\appendix

\section{An example of trajectory-independent microscopic work. \label{s.bgA}}

We exhibit a Hamiltonian $H(z,\lambda)$ and an evolution $\lambda(t)$ of the external parameter such that
\begin{equation}
	H(z(t_{\rm f}|z_0),\lambda(t_{\rm f})) - H(z_0,\lambda_0) = C
\,,
\label{e.A1}
\end{equation}
with $C$ independent of $z_0$\@.

Take the harmonic oscillator
\begin{equation}
	H(x,p,\lambda) = \frac{p^2}{2} + \frac{\omega^2 x^2}{2} -\omega^2 \lambda x
\,.
\label{e.A2}
\end{equation}
Call
\begin{eqnarray}
	\bar x(t|x_0,p_0) &=& x_0 \cos \omega t + p_0
	\frac{\sin \omega t}{\omega}\nonumber\\
	\bar p (t|x_0,p_0) &=& -\omega x_0 \sin\omega t + p_0 \cos
	\omega t
\label{e.A3}
\end{eqnarray}
the solution with $\lambda=0$\@.

For $\lambda(s)$ a function of time $s$, the solutions of the Hamiltonian equations starting from $(x_0,p_0)$ at $s=0$ are
\begin{eqnarray}
	x(t|x_0,p_0) &=&  \bar x(t) + \omega \int_0^t \lambda(s)
	\sin(\omega (t-s)) \ud s\\
	p (t|x_0,p_0) &=& \bar p(t) + \omega^2 \int_0^t \lambda(s)
	\cos(\omega(t-s)) \ud s
\label{e.A4}
\end{eqnarray}
with $\frac{\ud x}{\ud t} = p$ and $\frac{\ud p}{\ud t} - -\omega^2 x + \omega^2 \lambda(t)$\@. Assume that $\lambda_0=0$\@. Then
\bea
	H(x(t),p(t),\lambda(t)) &&- H(x_0,p_0,0) =
		-\omega^2\lambda(t) \bar x(t)
        + \omega^2 \bar p(t) \int_0^t \lambda(s) \cos \omega (t-s) \ud s +
        \nonumber \\ &&
		+\omega^3 \bar x(t) \int_0^t \lambda(s) \sin\omega(t-s) \ud s
        +\frac{1}{2} \omega^4 \left( \int_0^t \lambda(s) \cos
		\omega(t-s) \ud s \right)^2 +
        \nonumber \\ &&
        + \frac{1}{2}\omega^4 \left(\int_0^t \lambda(s) \sin
		\omega(t-s) \ud s \right)^2 - \omega^3 \lambda(t)\int_0^t
		\lambda(s) \sin\omega(t-s) \ud s.
\,.
\label{e.A5}
\eea
We can impose the condition that this quantity does not depend on $x_0$,$p_0$ provided that at time $t$
\begin{eqnarray}
	\int_{0}^t \lambda(s) \cos \omega (t-s) \ud s &=& 0\nonumber\\
	-\lambda(t) + \omega \int_0^t \lambda(s) \sin \omega(t-s) \ud s &=& 0
\,.
\label{e.A6}
\end{eqnarray}
Then using these two equalities, one has
\begin{equation}
	H(x(t),p(t),\lambda(t)) - H(x_0,p_0,0) =
	-\frac{\omega^2}{2}(\lambda(t))^2.
\label{e.A7}
\end{equation}
Thus if $\lambda(t)\ne0$, we can arrange that the microscopic work is independent of the initial condition and is non zero.

\section{An exactly solvable model\label{s.bgB}}

The system $A+B$ is formed of two two-levels atoms. The Hamiltonians of $A$ and $B$ are
\begin{equation}
	H_A = \left(
	\begin{array}{cc}
		0 & 0 \\
		0 & E_A
	\end{array}
	\right)
	\quad\hbox{and}\quad
	H_B = \left(
	\begin{array}{cc}
		0 & 0 \\
		0 & E_B
	\end{array}
	\right)
\,,
\label{eq:solv:ham}
\end{equation}
with eigenstates $|0_A\rangle$, $|+_A\rangle$, $|0_B\rangle$, $|+_B\rangle$, so that  the total Hamiltonian is in the basis $|0_A,0_B\rangle$, $|+_A,0_B\rangle$, $|0_A,+_B\rangle$, $|+_A,+_B\rangle$:
\begin{equation}
	H = \left( \begin{array}{cccc}
		0   & 0   & 0   & w         \\
		0   & E_A & 0   & 0         \\
        0   & 0   & E_B &  0        \\
		w^* & 0   & 0   & E_A + E_B
	\end{array}\right)
	\label{eq:solv:hammatrix}
\end{equation}
where $w$ is the interaction energy.

Calling $E_0 = E_A + E_B$, the eigenvalues of $H$ are
\begin{equation}
	\lambda_{\pm} = \frac{E_0 \pm \sqrt{E_0^2 + 4|w|^2}}{2}
\,.
\label{eq:solv:eva}
\end{equation}
as well as $E_A$ and $E_B$\@. The eigenstates of $E_A$ and $E_B$ are $|+_A,0_B\rangle$, $|0_A,+_B\rangle$, and the eigenstates of $\lambda_\pm$ are
\begin{equation}
	|\varphi_\pm\rangle = \frac{1}{N_\pm}\left( w|0_A,0_B\rangle
	+ \lambda_\pm |+_A,+_B\rangle \right)
\,,
\label{eq:solv:eve}
\end{equation}
so that
\begin{equation}
	|0_A,0_B\rangle = \frac{N_+ N_-}{w(\lambda_- -
	\lambda_+)}\left( \frac{\lambda_-}{N_-}|\varphi_+\rangle -
	\frac{\lambda_+}{N_+}|\varphi_-\rangle \right)
\label{e.bgB010}
\end{equation}
\begin{equation}
	|+_A+_B\rangle = \frac{N_+N_-}{\lambda_+ - \lambda_-}\left(
	\frac{1}{N_-}|\varphi_+\rangle -
	\frac{1}{N_+}|\varphi_-\rangle \right).
	\label{eq:solv:papb}
\end{equation}
Here $N_\pm = \sqrt{|w|^2 + |\lambda_\pm|^2}$ is the normalization
factor.

The initial state is $\rho_{A,\beta_A}\otimes
\rho_{B,\beta_B}$:
\bea
		\rho_{A,\beta_A}\otimes \rho_{B,\beta_B} &=&
		\frac{1}{Z_AZ_B}
      \Bigg(|0_A,0_B\rangle \langle 0_A,0_B| + e^{-\beta_A
		E_A}|+_A,0_B\rangle \langle +_A,0_B|
\nonumber \\ &&
		\quad +\, e^{-\beta_B E_B}|0_A,+_B\rangle \langle 0_A,+_B| +
		e^{-\beta_A E_A - \beta_B E_B}|+_A+_B\rangle
   \langle +_A+_B| 	\Bigg) \qquad
\,.
\label{eq:solv:ini}
\eea
Using these formulas one can compute
\begin{equation}
	\rho(t) = e^{-iH t}\rho_{A,\beta_A}\otimes \rho_{B,\beta_B}
	e^{iHt} = U(t) \rho_{A,\beta_A}\otimes \rho_{B,\beta_B} U(t)^+
	\label{eq:solv:evol}
\end{equation}
and verify that
\begin{equation}
	\tr\left( H_A \rho(t) \right) = \frac{E_A}{Z_AZ_B} \Bigg(
	e^{-\beta_AE_A} + e^{-\beta_AE_A - \beta_B
	E_B}\frac{|\lambda_+
	e^{-i\lambda_+t}-\lambda_-e^{-i\lambda_-t}|^2}{\left(
	\lambda_- - \lambda_+
	\right)^2} \, |w|^2 \;
    \frac{|e^{-i\lambda_+ t} -
	e^{-i\lambda_-t}|^2}{\left( \lambda_- - \lambda_+
	\right)^2}
	\Bigg)
\,.
\label{eq:solv:ena}
\end{equation}
Then
\begin{equation}
	S(\rho(t)|\rho_{A,\beta_A}\otimes \rho_{B,\beta_B}) =
	\tr\left( (\beta_A H_A + \beta_B H_B) (\rho(t) -
	\rho_{A,\beta_A}\otimes \rho_{B,\beta_B}) \right)
\label{eq:solv:kl}
\end{equation}
\begin{equation}
	\delta E_V(\rho) = -\tr\left( (H_A + H_B)(\rho(t) -
	        \rho_{A,\beta_A}\otimes \rho_{B,\beta_B}) \right)
\label{eq:solv:int}
\end{equation}
and
\begin{equation}
	S(\rho(t)|\rho_{A,\beta_A}\otimes \rho_{B,\beta_B}) =
	\frac{\beta_A E_A + \beta_B E_B}{Z_AZ_B} \left( 1-
	e^{-\beta_A E_A - \beta_A E_B}
	\right)
        \frac{|w|^2 \; \sin^2\frac{(\lambda_+ - \lambda_-)t}{2}}
            {\left( \frac{\lambda_+ - \lambda_-}{2}
	\right)^2}
\label{e.bgB.1}
\end{equation}
\begin{equation}
	\du E_V(\rho) = - \frac{E_A + E_B}{Z_AZ_B}
     \left( 1-  e^{-\beta_A E_A - \beta_A E_B}	\right)
    |w|^2 \;
    \frac{ \sin^2\left(\left(\lambda_+ - \lambda_-\right)t\right)  }
    { \left( \lambda_- - \lambda_+ 	\right)^2 }
\,.
\label{e.bgB.2}
\end{equation}
Using \Eqref{e.bg6.13}, one obtains
\begin{equation}
	\du E_A (\rho) = \frac{E_A}{Z_AZ_B}\left( 1-
	e^{-\beta_A E_A - \beta_A E_B}
	\right) |w|^2 \;
	\frac{\sin^2\left(\left(\lambda_+  - \lambda_- \right) t\right)}
    {\left(	\lambda_- -	\lambda_+	\right)^2 }
\,.
\label{e.bgB.3}
\end{equation}
Here these quantities are periodic functions of period $\frac{2\pi}{\lambda_+-\lambda_-} = \frac{2\pi}{\sqrt{E_0^2 +4|w|^2}}$\@. Near resonance, where $\lambda_+ \simeq \lambda_-$, $w\simeq 0$, $E_0 = E_A + E_B \simeq 0$ and we recover that $\du E_A(\rho) \simeq K(\beta_A - \beta_B)t$ from \Eqref{e.bgB.3}.

\section{Example: forced harmonic oscillator\label{s.bgC}}

We take the Hamiltonian
\begin{equation}
	H = -\frac{1}{2}\frac{\ud^2}{\ud x^2} + \frac{\omega^2 x^2}{2}
	+ \lambda(t) x = H_0 + \lambda(t) x
\,,
\label{eq:ex:hamilt}
\end{equation}
with the condition $\lambda(0)=0$\@. The classical action is
\bea
S(x,t|x') &=& \frac{\omega}{2\sin \omega t}
   \biggl( (x^2 +
	x'^2)\cos \omega t - 2 x x' - \frac{2x}{\omega} \int_0^t
	\lambda(s) \sin \omega s \ud s  \biggr.
\nonumber\\&& \qquad\qquad \biggl.
    - \frac{2x'}{\omega}\int_0^t
	\lambda(s) \sin \omega(t-s) \ud s + C(t)
	\biggr)
\,,
\label{e.bgC.1}
\eea
where $C(t)$ does not depend on $x$ or $x'$\@. The quantum propagator is
\begin{equation}
	G(x,t|x',0) \simeq \exp(iS(x,t|x',0))
\,,
\label{e.bgC2}
\end{equation}
where ``$\simeq$'' indicates that we have not written the normalization factor. This factor does not depend on $x$ or $x'$ and is at the moment unimportant. The thermal state for $\lambda=0$ is
\begin{equation}
	\rho_\beta \simeq \exp\left( -\frac{i\omega}{2\sin(i\omega
	\beta)} \left( (y^2 + y'^2) \cos i\omega\beta - 2y y' \right) \right)
\,.
\label{e.bgC.3}
\end{equation}
The time-evolved state at time-$t$ is
\begin{equation}
	\rho(t,x|x') = \iint G(x,t|y)\rho_\beta(y|y')G(y't|x')^*\ud y
	\ud y'
\,.
\label{e.bgC.4}
\end{equation}
The energy at time-$t$, using $\lambda(t)=0$, is
\begin{equation}
	E(t) = \int \ud x \left. H_{0,x}\rho(t,x|x')
	\right|_{x'=x}
\,,
\label{eq:ex:e}
\end{equation}
with
$H_{0,x} = -\frac{1}{2}\frac{\ud^2}{\ud x^2} + \frac{\omega^2 x^2}{2}$\@. Define
\begin{eqnarray}
	I_1 &=& \int_0^y \lambda(s)\frac{\sin\omega s}{\sin \omega t}\ud s \\
	I_2 &=& \int_0^t \lambda(s) \frac{\sin(\omega(t-s)}{\sin\omega t}\ud s\\
	A &=& -i I_1 + i I_2\left( \frac{\sin\omega t}{\sin i\omega	\beta} - \frac{\sin\left( \omega t + i\omega \beta
    	\right)}{\sin i\omega \beta} \right)\\
	A' &=& iI_1 + iI_2\left( \frac{\sin\omega t}{\sin i\omega	\beta} + \frac{\sin\left( i\omega \beta-\omega t
    	\right)}{\sin i\omega \beta}\right)
\,.
\label{eq:ex:define}
\end{eqnarray}
The calculation of the double Gaussian integral in \Eqref{e.bgC.4} gives
\begin{equation}
	\rho(t,x|x') = \frac{1}{N(t)}\exp\left(
	-\frac{\omega}{2}(x^2+x'^2)\coth \omega \beta +
	\frac{\omega x x'}{\sinh \omega \beta} + Ax + A'x' \right)
\,,
\label{eq:ex:doublegauss}
\end{equation}
where $N(t)$ is the normalization factor
\begin{equation}
	N(t) = \exp\left( \frac{1}{4}\frac{(A+A')^2}{\omega \coth
	\omega \beta - \frac{\omega}{\sinh \omega \beta}}
	\right)\frac{\sqrt{2\pi}}{\sqrt{2(\omega \coth
	\omega\beta-\frac{\omega}{\sinh \omega \beta})}}
\,.
\label{eq:ex:norm}
\end{equation}
The action of the Hamiltonian on the propagated state is
\begin{equation}
	H_{0,x}\rho(t,x|x') = \left( \frac{1}{2}\omega \coth \omega
	\beta - \frac{1}{2}\left( (-\omega x\coth
	\omega\beta+\frac{\omega x'}{\sinh \omega \beta})+A \right)^2
	+ \frac{\omega^2 x^2}{2}\right) \rho(t,x|x')
\,.
\label{eq:ex:h0x}
\end{equation}
We define the variable $X$ as
\begin{equation}
	X = x- \frac{A+A'}{2\left(-\omega \coth
	\omega\beta+\frac{\omega }{\sinh \omega \beta}\right)}
\,.
\label{eq:ex:chvar}
\end{equation}
Then the energy of the propagated state at time $t$ is
\bea
	E(t) &=&
    \frac{\sqrt{2\left( \omega \coth \omega \beta - \frac{\omega}{\sinh \omega \beta}
		\right)}}{\sqrt{2\pi}}
\nonumber\\ &&
\int \ud X \exp\left( -\left(
		\omega \coth \omega \beta - \frac{\omega}{\sinh \omega
		\beta} \right) X^2
		\right)
\nonumber\\ &&\;
        \times
		\Bigg[
		\frac{1}{2}\omega \coth \omega \beta -
		\frac{1}{2}
		\left(
		X\left(
		- \omega \coth \omega \beta +
		\frac{\omega}{\sinh \omega \beta}
		\right)
		+ \frac{1}{2}(A - A')^2\right)
\nonumber \\ &&  \qquad\quad
		+\frac{\omega^2}{2}
		\left(
		X + \frac{1}{2} \frac{A+A'}{\omega \coth \omega \beta
		- \frac{\omega}{\sinh \omega \beta}}
		\right)^2
		\Bigg]
\,,
\label{eq:ex:et}
\eea
and $E(0)$ is the value of $E(t)$ at $t=0$, so that
\begin{equation}
	E(t) - E(0) = \frac{1}{8}\left( \frac{(A+A')^2}{\left(\coth \omega
	\beta- \frac{1}{\sinh \omega \beta}\right)^2}- (A-A')^2 \right)
\,.
\label{eq:ex:et2}
\end{equation}
Finally using the values of $A$ and $A'$ in terms of $I_1$ and $I_2$, we obtain
\begin{equation}
	E(t) - E(0) = \frac{1}{2}\left( I_1^2 + I_2^2 + 2I_1 I_2 \cos
	\omega t\right)
\,.
\label{eq:ex:et3}
\end{equation}
This is independent of $\beta$ and is positive. As a corollary, this result is valid if one propagates any eigenstate of the Hamiltonian $H_0$\@. One can also derive the classical energy
\begin{equation}
	E(t) - E(0) = \left\langle
	H(x(t|x_0,p_0),p(t|x_0,p_0),\lambda=0) -
	H(x_0,p_0,\lambda=0)\right\rangle_{\rho_{\beta}(\lambda=0)}
\,,
\label{eq:ex:et4}
\end{equation}
where $\rho_\beta(\lambda=0)$ is the classical thermal state. One uses the equations of motion
\begin{eqnarray}
	x(t|x_0,p_0) &=& x_0 \cos\omega t + p_0 \frac{\sin \omega
	t}{\omega} + \omega \int_0^t \lambda(s) \sin\omega(t-s) \,\ud s
	\\
	p(t|x_0,p_0) &=& -\omega x_0 \sin \omega t + p_0 \cos \omega t
	+ \omega^2 \int_0^t \lambda(s) \cos \omega (t-s)\, \ud s
\label{eq:ex:xp}
\end{eqnarray}
\begin{equation}
	\rho_\beta(\lambda=0) = \frac{1}{N_\beta}\exp\left(
	-\beta\left( \frac{p^2}{2} + \frac{\omega^2 x^2}{2} \right)
	\right)
\,,
\label{eq:ex:therm0}
\end{equation}
and then
\begin{equation}
	E(t) - E(0) = \frac{1}{2}\omega^4 \left( \left( \int_0^t
	\lambda(s) \cos \omega(t-s)\, \ud s
	\right)^2 + \left( \int_0^t \lambda(s)\sin\omega(t-s)\,\ud s
	\right)^2 \right)
\,.
\label{eq:ex:et5}
\end{equation}
If $\lambda(0) = 0$ but $\lambda(t) \ne 0$, one gets
\bea
		E(t) - E(0) &=& \frac{1}{2}\omega^4 \Bigg( \left( \int_0^t
		\lambda(s) \cos \omega(t-s)\, \ud s
		\right)^2 + \left( \int_0^t
		\lambda(s)\sin\omega(t-s)\,\ud s
		\right)^2
\nonumber\\ && \qquad
      - \frac{2\lambda(t)}{\omega} \int_0^t
		\lambda(s)\sin\omega(t-s)\, \ud s \Bigg)
\,.
\label{eq:ex:et6}
\eea
This can be negative, for example if $\lambda(t) = t$:
\begin{equation}
	E(t) - E(0) = 1 - \frac{t^2 \omega^2}{2} - \cos \omega t <0
\,.
\label{eq:ex:et7}
\end{equation}


\begin{thebibliography}{20}
\bibitem[{not({\natexlab{a}})}]{note:limited}
\bibinfo{note}{In \cite{master} and \cite{framework} we found the relative
  entropy to be related to dissipation, but we there dealt with stochastic
  dynamics and our conclusions were limited to short times.}
\bibitem[{not({\natexlab{b}})}]{note:suppressinverse}
\bibinfo{note}{Henceforth we suppress the qualifier``inverse'' in referring to
  various $\beta$'s as temperatures.}
\end{thebibliography}

\end{document}